\documentclass[]{aa}
\usepackage{graphicx}
\usepackage{psfig}
\newcommand{\kms}  {\mbox{\rm km\,s$^{-1}$}}
\newcommand{\cms}  {\mbox{\rm cm\,s$^{-1}$}}
\newcommand{\mA} {\mbox {m\AA}}

\newcommand{\teff}  {\mbox{${\rm T}_{\rm eff}$}\,}

\newcommand{\logg}  {\mbox{{\rm log}\,$g$}}

\newcommand{\LiI} {\ion{Li}{i}}

\newcommand{\ld} {\ldots}

\begin{document}

\title{The lithium content of the Galactic Halo stars}

\author{C.~Charbonnel \inst{1,2} and F.~Primas \inst{3,2}}

\offprints{C.~Charbonnel}

\institute{
Geneva Observatory, 51, chemin des Maillettes, CH--1290 Sauverny,
Switzerland\\
\email{Corinne.Charbonnel@obs.unige.ch}
\and
Laboratoire d'Astrophysique de Toulouse et de Tarbes, CNRS UMR 5572, 14 av. E.
Belin, F--31400 Toulouse, France\\ 
\and 
European Southern Observatory, Karl-Schwarzschild Str. 2,
D--85748 Garching b. M\"{u}nchen, Germany\\
\email{fprimas@eso.org}
\thanks{
Visiting Astronomer at the LATT - OMP, Toulouse, France}
}

\date{Accepted for publication}

\titlerunning{The Lithium Content of the Galactic Halo Stars}
\authorrunning{Charbonnel \& Primas}

\abstract{
Thanks to the accurate determination of the baryon density of the universe
by the recent cosmic microwave background experiments, updated predictions
of the standard model of Big Bang nucleosynthesis now yield the initial
abundance of the primordial light elements with unprecedented precision.
In the case of $^7$Li, the CMB+SBBN value is significantly higher than the
generally reported abundances for Pop II stars along the so-called Spite
plateau. In view of the crucial importance of this disagreement, which has
cosmological, galactic and stellar implications, we decided to tackle the
most critical issues of the problem by revisiting a large sample of
literature Li data in halo stars that we assembled following some strict
selection criteria on the quality of the original analyses.

In the first part of the paper we focus on the systematic uncertainties
affecting the determination of the Li abundances, one of our main goal 
being to look for the ``highest observational accuracy achievable" for
one of the largest sets of Li abundances ever assembled. We explore in
great detail the temperature scale issue with a special emphasis on reddening.
We derive four sets of effective temperatures by applying the same
colour-\teff\,calibration but making four different assumptions about
reddening and determine the LTE lithium values for each of them. We compute
the NLTE corrections and apply them to the LTE lithium abundances.
We then focus on our ``best" (i.e. most consistent) set of temperatures 
in order to discuss the inferred mean Li value and dispersion in several
\teff and metallicity intervals. The resulting mean Li values along the
plateau for [Fe/H] $\leq$ -1.5 are 
$A(Li)_{NLTE}$ = 2.214$\pm$0.093 and 2.224$\pm$0.075 when the lowest 
effective temperature considered is taken equal to 5700~K and 6000~K
respectively. This is a factor of $\sim$~2.48 to 2.81
(depending on the adopted SBBN model and on the 
effective temperature range chosen to delimit the plateau)
lower than the CMB+SBBN determination. We find no evidence of intrinsic
dispersion. 
Assuming the correctness of the CMB+SBBN prediction, we are
then left with the conclusion that the Li abundance along the plateau is
not the pristine one, but that halo stars have undergone surface depletion
during their evolution. 

In the second part of the paper we further dissect our sample in search
of new constraints on Li depletion in halo stars. By means of the Hipparcos 
parallaxes, we derive the evolutionary status of each of our sample stars,
and re-discuss our derived Li abundances. A very surprising result emerges
for the first time from this examination. 
Namely, the mean Li value as well as the dispersion 
appear to be lower (although fully compatible within the errors) for the
dwarfs than for the turnoff and subgiant stars. 
For our most homogeneous dwarfs-only sample with [Fe/H]$\leq$-1.5, 
the mean Li abundances are 
$A(Li)_{NLTE}$ = 2.177$\pm$0.071 and 2.215$\pm$0.074 when the lowest 
effective temperature considered is taken equal to 5700~K  
and 6000~K respectively. This is a factor of 2.52 to 3.06 
(depending on the selected range in \teff for the plateau and on the SBBN 
predictions we compare to) lower than the CMB+SBBN primordial value. 
Instead, for the post-main sequence stars the corresponding 
values are 2.260$\pm$0.1 and 2.235$\pm$0.077, which correspond to a 
depletion factor of 2.28 to 2.52.

These results, together with the finding that all the stars with Li
abnormalities (strong deficiency or high content) lie on or originate
from the hot side of the plateau, lead us to suggest that the most massive
of the halo stars have had a slightly different Li history than their less
massive contemporaries. In turn, this puts strong new constraints on the
possible depletion mechanisms and reinforces Li as
a stellar tomographer.

}

\maketitle

\section{The Lithium Paradigm}

From the time of its discovery in the stellar atmospheres of very metal-poor
Population II stars (Spite \& Spite 1982a), lithium has been considered a key
diagnostic to test and constrain our understanding and description of the
primordial Universe, of stellar interiors and evolution, and of spallation
physics (for the latter two especially so if combined with abundances of
beryllium and boron). 

Lithium-7 is one of the four primordial isotopes that have been formed 
in observable quantities by nuclear reactions during the first minutes 
of the Universe (e.g. Olive, Steigman, \& Walker 2000 and references therein).
Together with deuterium, helium-3 and helium-4, knowledge of its primordial 
abundance provides one of the main observational constraints on the 
baryon-to-photon ratio $\eta \propto \Omega_b h^2$ which is the only 
free parameter of the standard Big Bang Nucleosynthesis (SBBN). 

Among these light elements, lithium is one of the easiest to observe
(its resonant doublet falls at 670\ nm, i.e. easily accessible from ground
telescopes even of small sizes), which explains the wealth of data available
in the literature. 

Another reason for such a large literature database is connected to the
important finding of a remarkably flat and constant Li abundance among
Galactic halo dwarf stars spanning a wide range of effective temperatures
and metallicities (the so-called Spite plateau, cf Spite \& Spite 1982a,b). 
This result came as a surprise. At that time indeed it was generally believed
that the primordial A(Li)\footnote{A(Li) = 12 + log N(Li)/N(H)} abundance
was in the range 3.0--3.3, which corresponds to the value measured in
meteorites and also to the maximum value detected in Population\,I stars,
both in the field and in open clusters (see the review by Boesgaard \&
Steigman 1985). If this were the case, then the ``constant" but lower value
of lithium along the Spite plateau would have required that all the oldest
stars of our Galaxy had suffered a uniform depletion by about a factor of
10. Although several mechanisms could be conjectured to modify the surface
lithium abundance (proto-stellar destruction, microscopic diffusion,
turbulent diffusion, mass loss), they were also suspected to depend on the
stellar mass (i.e., on the effective temperature). Consequently the
constancy of the Li plateau was used (and is actually still used very
often) as an argument to say that these processes were in fact not efficient
in Pop II stars.

The other interpretation was then that the plateau value represents the
amount of Li
produced during the Big Bang, and that the Galaxy had been enriched in its
Li content by a factor of at least 10 since its birth\footnote{Actually the 
primordial Li abundance could even be lower than the plateau value because 
of production in the early Galaxy (Ryan et al. 1999; Suzuki et al. 2000).}. 
The fact that lithium is produced in several other nucleosynthetic sites (i.e.
$\alpha-\alpha$ fusion, spallation reactions, late stellar evolutionary
stages, like AGB stars, novae, etc.; see Romano et al. 2003 and references
therein), none of which has been quantitatively and accurately estimated
nor strongly constrained by observations, complicates the final
interpretation of its Galactic evolution. 

Recent results on cosmic microwave background anisotropies, most particularly 
from the {\sl Wilkinson Microwave Anisotropy Probe (WMAP)} experiment 
(Bennet et al. 2003; Spergel et al. 2003) allowed an unprecedented precision 
on the determination of the baryon-to-photon ratio $\eta$ and revealed that
Li seems to lie between the two extreme
solutions discussed above. The {\sl WMAP} data alone lead to $\Omega_b h^2 = 
0.0237 \pm 0.001$, or $\eta = 6.5^{+0.4}_{-0.3} \times 10^{-10}$.
When combined with additional CMB experiments (CBI, Pearson et al. 2003; 
ACBAR, Kuo et al. 2002) and with measurements of the power spectrum (2dF
Galaxy Redshift Survey, Percival et al. 2001; Ly$\alpha$ forest, Croft et
al. 2002, Gnedin \& Hamilton 2002), the resulting 
values are 
$\Omega_b h^2 = 0.0224 \pm 0.0009$ or $\eta = 6.1^{+0.3}_{-0.2} \times 10^{-10}$.
With this value of $\eta$, updated SBBN predictions now allow a precise
determination of the primordial abundances of the light elements D, $^3$He, 
$^4$He and $^7$Li that we can compare with observations in low-metallicity 
environments. The {\sl WMAP}+SBBN determinations of these abundances in the
most two recent studies (Coc et al. 2004, Cyburt 2004, Serpico et al. 2004) are summarised in
Table 1. 

\begin{table*}
\caption{{\sl WMAP}+SBBN primordial abundances predicted when 
$\eta = (6.14 \pm 0.25) \times 10^{-10}$ (or $\Omega_b h^2 = 0.0224 \pm$0.0009; 
Spergel et al. 2003) is adopted}
\begin{center}
\begin{tabular}{cccc}
\hline
\hline
\multicolumn{1}{c}{}
& \multicolumn{1}{c}{Coc et al. (2004)} 
& \multicolumn{1}{c}{Cyburt (2004)} 
& \multicolumn{1}{c}{Serpico et al. (2004)} \\
\hline
D/H & (2.60$^{+0.19}_{-0.17}) \times 10^{-5}$ & (2.55$^{+0.21}_{-0.20}) \times 10^{-5}$ 
& (2.58$^{+0.19}_{-0.16}) \times 10^{-5}$ \\
Y$_P$ & 0.2479$\pm$0.0004 & 0.2485$\pm$0.0005 & 0.2479$\pm$0.0004 \\
$^3$He/H & (1.04$\pm 0.04)\times 10^{-5}$ & (1.01$\pm 0.07) \times 10^{-5}$ 
& 1.03$\pm$0.03 \\.
$^7$Li/H & (4.15$^{+0.49}_{-0.45}) \times 10^{-10}$ & (4.26$^{+0.91}_{-0.86}\times) 10^{-10}$ 
& (4.6$^{+0.4}_{-0.4}) \times 10^{-10}$\\
\hline
\end{tabular}
\end{center}
\end{table*}

A very good agreement is achieved between the primordial abundance of
deuterium derived from {\sl WMAP}+SBBN and the average value of D/H
observations in cosmological clouds along the line of sight of quasars
(Kirkman et al. 2003). On the other hand the observational data of $^3$He
in galactic HII regions are scarce and must be corrected for contamination
of the observed gas by ejecta from earlier generations of stars (e.g. Tosi
1998, Charbonnel 2002). The upper limit to the primordial abundance
recommended by Bania et al. (2002) is however quite consistent with the
CMB-derived value. 
Finally the CMB-predicted primordial $^4$He abundance is higher than the
values derived from the determinations in complex low-metallicity HII regions
(both galactic and extra-galactic) and the extrapolation to zero oxygen
abundance (Izotov \& Thuan 2004, Olive \& Skillman 2004 and references
therein). However the difference is relatively modest (2-3$\%$) and it may
simply call for further exploration of the systematic effects in the
abundance analysis. 

The most critical case concerns $^7$Li, the CMB-derived primordial abundance
of which is clearly higher (by about a factor of 3) than the current
determinations in low-metallicity halo stars. 
This result seems to be very robust with respect to the nuclear uncertainties 
on the SBBN reactions although Coc et al. (2004) show that the discrepancy 
could be resolved by an increase of a factor of $\sim$ 100 of the
$^7$Be({\it d,p})2$^4$He reaction rate. Although this is not supported by the
data currently available, this issue has to be further investigated
experimentally. Should this nuclear solution be excluded, we would then be
left with the astrophysical solution. 
Namely, with the conclusion that the Li abundance that we see at the surface
of halo stars is not the pristine one, but that these stars have undergone
surface lithium depletion at some point during their evolution. This
possibility has been discussed many times in the literature. Several physical
mechanisms have been invoked, but all the current models encounter
considerable difficulties to reconcile a non negligible
depletion of lithium with both the flatness and the small dispersion along 
the so-called Li plateau (see the review by Pinsonneault et al. 2000 and
Talon \& Charbonnel 2004 for more recent references). 
The challenge thus still remains to identify the process (or processes) by
which a reduction by a factor of $\sim$3 could occur so uniformly in stars
over a large range in effective temperature and metallicity. 

With the CMB constraint, we are now entering a golden age for Li as both a 
baryometer and a stellar tomographer\footnote{Since lithium is destroyed 
at quite low temperatures (for stellar interiors) of the order of $2.5
\times 10^6\,K$, it is a powerful tool to identify the mechanisms active
in stellar interiors and responsible for convective and/or radiative
transports, mixing, diffusion, presence of gravitational waves. Together
with beryllium and boron, that burn at $3.5 \times 10^6\,K$ and $5.0 \times
10^6\,K$ respectively, lithium abundances allow us to make a stellar
tomography of the external atmospheric layers where these three light
nuclides are ``nuclearly" preserved (since the epoch of formation when
looking at unevolved objects).}. 
In this quest however one has still to pay special attention to the 
observational analysis and determination of the lithium abundances in
the most metal-poor, thus the oldest stars of our Galaxy. As a matter of
fact and despite the large number of spectroscopic data that has become
available in the last two decades, there are still on-going debates on
the patterns of the plateau, like its thinness, the possible existence
of a spread and of a dependence of A(Li) with metallicity and effective
temperature. These characteristics must be precisely determined in order
to constrain the physical processes which lead to Li depletion in Pop II
stars as well as those of Galactic production. 

\section{The ``Lithium Plateau Debate"?}

Deliyannis et al. (1993) were among the first to present 
evidence for the existence of dispersion (of the order of $\pm$20\%
about the mean, derived from the ``equivalent width-colour'' plane), followed
by Thorburn (1994) and Norris et al. (1994), the latter being the first to
also have found a dependence of A(Li) on both \teff\, and metallicity. 
Molaro et al. (1995) counter-argued these findings showing that when a fully
consistently determined temperature scale is used (in their case, the Fuhrmann
et al. 1994 scale, derived from Balmer lines fitting), no dispersion nor
tilt is found (Li abundances are mostly sensitive to \teff), but they were
shortly followed by Ryan et al. (1996) who once again confirmed the slopes.
They argued that the Molaro et al. sample was plagued by the inclusion of
subgiants that may have affected their final outcome. On the intrinsic scatter
issue, they noted that there were some stars, characterized by very similar
parameters (colour, \teff\ and metallicity), but that turned out from
multiple measurements to have very different Li abundances. The debate kept
being very alive: Spite et al. (1996) explored further the \teff\ scale
issue (comparing different \teff\,determinations) and found that the rms
scatter of the Li abundance was between 0.06 and 0.08\,dex, hence very small
if real. Bonifacio \& Molaro (1997) re-selected their sample, this time
excluding possible outliers (like the abovementioned subgiants) and once
again came to the conclusion of no intrinsic dispersion nor dependence of
A(Li) on metallicity, but of a tiny trend with the temperature. They concluded
that the finding or not of a trend with effective temperature may well depend
on the adopted \teff\ scale. 

It is only towards the end of the 90s when a
full agreement on the absence of intrinsic dispersion was reached: Ryan et
al. (1999), by analysing a new sample of 23 stars covering a narrow range
in \teff\,(6050-6350~K) and in metallicity ($-$3.5 -- $-$2.5), claimed
that the intrinsic spread is effectively zero, i.e. 0.031\,dex, at the
1$\sigma$ level (to be compared to their formal errors of 0.033\,dex).
However, they still recovered the dependence on metallicity, at the level
of dA(Li)/d[Fe/H] = 0.118$\pm$0.023~(1$\sigma$) dex per dex, i.e. very
similar to the slopes previously found. The trend with \teff\,, if any, is
likely to be meaningless because of the very narrow \teff\, range there
explored. 

This is when we started developing our project. By comparing the data samples
analysed by different authors (starting with the Bonifacio \& Molaro 1997
and Ryan et al. 1996, 1999 because of their final, opposite claims)
we noticed that for some stars, common to several analyses, very discrepant
Li abundances were reported, which could have clearly influenced some of the
early claims for dispersion, and they could still play a role in the current
debate about the existence of a slope between Li and \teff\, and [Fe/H]. 

In order to further tackle these issues, we decided to re-analyse the large
sample of Li abundances available in the literature (\S~3 and 4) from a
different perspective. One of our main goals is to focus on the systematic
uncertainties affecting the determination of Li abundances. First, because
the Li abundance is strongly dependent on the assumed temperature, we explore
further the temperature scale issue (\S~5) with the aim of deriving what
the best achievable accuracy may be for a temperature scale derived in a
consistent manner. We put special emphasis on reddening, usually an
underestimated source of error. 
We derive four sets of effective temperatures by applying the same
colour-T$_{\rm eff}$ calibration but making four different assumptions about 
reddening.
We then derive the LTE lithium abundances for each of these sets and compute 
the NLTE corrections. Then we select our ``best" (in terms of consistency)
set of temperatures in order to  
determine the mean Li abundance and the dispersion for one of the largest
sample of halo stars ever studied in a consistent way 
(\S~6 and 7). This allows us to derive preliminary results on the mean
lithium abundance and dispersion which can be compared to previous analyses 
(\S~8).

Secondly we look afresh at our Li abundances together with the evolutionary
status of each target (\S~9) in order to get clues on the internal processes
that may have been involved in modifying the Li abundances along the Plateau.
This is why we did
not restrict a priori our sample to any specific evolutionary status. 
We then discuss the lithium abundance along the plateau for the dwarf stars
only (\S~10) and look at its behaviour in subgiants (\S~11). We test whether
our results on the dispersion and trends can be altered by the presence of
binary stars in the sample (\S~12), and we finally inspect the cases of stars
with extreme lithium abundances (\S~13). Then we discuss the current status
of Pop II stellar models in view of our observational results (\S~14).
Finally we summarise our results and conclude on some remaining open
questions (\S~15).

\section{Sample selection: A Critical analysis of the literature}

The database of Li abundances measured in Galactic stars and available in the
literature is huge. During this work, we restricted our search
to the main observational analyses from the early 90 onwards. 

\begin{table}[b]
\label{table2}
\tiny{
\parbox{90mm}{
\caption[]{Literature sources} 
\begin{flushleft}
\begin{tabular}{lcrcc}
\noalign{\smallskip}
\hline
\hline
\noalign{\smallskip}
 Authors   & R & S/N & T/g/Fe$^{\mathrm{a}}$ & Note$^{\mathrm{b}}$ \\ 
\noalign{\smallskip}
\hline
1. Ryan et al. 1996 & 38,000 & $>$100 &  p/l/l & O,T \\
2. Ryan et al. 1999 & 40,000 & $>$100 &  p/l/l & T \\
3. Ryan et al. 2001a,b & 50,000 & $>$100 &  p/l/l & O,T \\
4. Bonifacio \& Molaro 1997 & \ldots & \ldots & p/l/l & T \\
5. Pilachowski et al. 1993 & 30,000 & $>$100 & p/p/s & O \\
6. Hobbs \& Thorburn 1991 & 30,000 & $>$100 & l/l/l & O,T \\
7. Thorburn 1994 & 28,000 & $>$80 & p/l/l & O \\
8. Molaro et al. 1995 & \ldots & \ldots & p/l/l & T \\
9. Spite et al. 1996 & \ldots & $>$80 & p/p/s & T \\
10. Ryan \& Deliyannis 1998 & 42,000 & $>$70 & p/p/s & O \\ 
11. Gutierrez et al. 1999 & 30,000 & $>$100 & \ldots & O \\
12. Fulbright 2000 & 50,000 & $>$100 & s/s/s & O \\
13. Ford et al. 2002 & 50,000 & $>$150 & l/l/l & O \\ 
\noalign{\smallskip}
\hline
\end{tabular}
\end{flushleft}
\begin{list}{}{}
\item[$^{\mathrm{a}}$] \teff/\logg/[Fe/H]: l=literature; s=spectroscopy;
p=photometry
\item[$^{\mathrm{b}}$] O=new set of observations; T=new \teff\,scale (with
Li EW collected from literature) 
\end{list}{}{} 
}
}
\end{table}

\begin{table*}[t]
\label{table3}
\caption{The data sample and its main characteristics, as found in the literature. 
The whole table is available on-line}
\footnotesize{
\parbox{220mm}{
\setlength{\tabcolsep}{0.15cm}
\begin{flushleft}
\begin{tabular}{rrrcrccccc}
\noalign{\smallskip}
\hline
\hline
\noalign{\smallskip}
 HIP  & HD   & BD/CD      & G       & V      & \teff$_{Lit}$  & \logg$_{Lit}$ & [Fe/H]$_{Lit}$ & EW$_{Lit}$ & Ref$^{\mathrm{a}}$ \\  
      &      &            &         & mag    & K              & \cms          & dex            & m\AA       &                    \\
\noalign{\smallskip}
\hline
  484 &   97 & $-$20 6718 &         &  9.660 & 5000           & \ld & $-$1.23              & 12.1           & 5    \\
  911 &      &            & 266-060 & 11.800 & 5890           & 2.2 & $-$1.84              & 30.8           & 3    \\ 
 2413 & 2665 & $+$56 70   &         &  7.729 & 5050\ldots5100 & 3.6 & $-$1.80\ldots$-$1.89 & 15.0\ldots20.0 & 5,12 \\
 3026 & 3567 & $-$09 122B & 270-023 &  9.252 & 5858\ldots5930 & 3.7 & $-$1.20\ldots$-$1.34 & 45.0           & 1,4  \\ 
 3430 &      & $+$71 31   & 242-065 & 10.202 & 6026\ldots6170 & \ld & $-$1.91\ldots$-$2.20 & 31.0           & 1,4  \\ 
 \ld  & \ld  & \ld        & \ld     & \ld    & \ld            & \ld & \ld                  & \ld            & \ld  \\
\noalign{\smallskip}
\hline
\end{tabular}
\end{flushleft}
\begin{list}{}{}
\item[$^{\mathrm{a}}$] The references are those given in Table 2
\end{list}{}{} 
}
}
\end{table*}

\begin{figure}[t]
\rotatebox{0}{\resizebox{9cm}{!}{\includegraphics{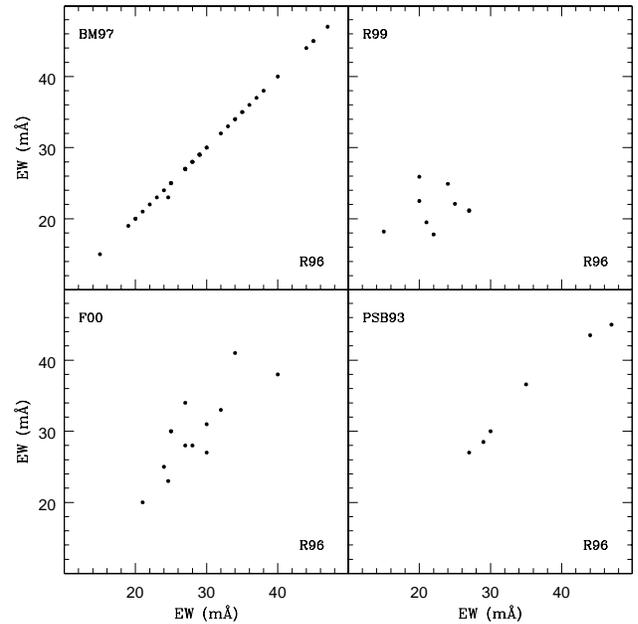}}}
\caption{The comparison of EWs as measured/adopted by different authors for
stars in common. The literature sources used for these comparisons are
reported in the upper left and lower right corners for the {\it y-} and
{\it x-axis} respectively.}
\label{logg}
\end{figure}

We assembled our final data sample from 13 literature sources (cf Table~2).
The main ones were the works by Ryan et al. (1996, 1999, hereafter
R96, R99), Bonifacio \& Molaro (1997, hereafter BM97), and Pilachowski et
al. (1993, hereafter PSB93), the last one in order to include a large set
of subgiant stars. R96, R99, and PSB93 analysed newly observed spectra and
derived new temperatures, whereas BM97 collected a large sample of stars for
which the equivalent width of the Li\,I line had already been measured and
re-computed the Li abundance based on a new temperature scale. This first
list of targets was then
complemented by objects taken from Hobbs \& Thorburn (1991), Thorburn (1994,
for those very few stars which had not been already re-analysed by Ryan et
al. 1996), Molaro et al. (1995), Spite et al. (1996), Ryan \& Deliyannis
(1998), Gutierrez et al. (1999), Fulbright (2000), Ryan et al. (2001a,b),
and Ford et al. (2002). We gave our preference to original works, i.e.
works that added new observations or that re-analysed literature values based
on a new temperature scale. This exercise left us with a sample of 146 stars,
covering the metallicity range between [Fe/H]~=~$-$1.0 and $-$3.5, the
temperature interval \teff~=~4500--6500\,K, and the surface gravity range
between \logg~=~3.0--5.0. In other words, we are sampling the main sequence,
subgiant and giant evolutionary stages. Table~3\footnote{available in its
entirety on-line} presents the data sample and its main characteristics
in terms of nomenclature (including cross-identifications), stellar parameters
and the equivalent widths of the \LiI\, line as found in the literature. For
those objects, for which multiple determinations are available, the minimum
and maximum values are listed. For a more detailed comparison, Figure~1
shows how the equivalent widths measured (used) by some of the literature
sources listed in Table~1 compare to each other. For the purpose of this test,
we selected those works that had the largest number of stars in common.

For completion, we note that there have been three other recent works that
have also made use of the large database of Li measurements available from
the literature. Two of them had different scientific goals and they both
used (after a critical selection) the Li abundances as found in the
literature: Romano et al. (1999) re-assessed the Galactic evolution of
lithium, whereas Pinsonneault et al. (2002) compared the most recent Li
abundances to theoretical predictions of models including rotational mixing
and examined them for trends with metallicity. This is why they do not appear
in our list of literature sources. The third, most recent and most similar
work to ours is the one from Mel\'endez \& Ram\'{\i}rez (2004), who studied
the behavior of the A(Li) plateau and its trends in a sample of 41 dwarf
stars. An improved InfraRed Flux Method-based temperature scale was derived
(Ram\'{\i}rez \& Mel\'endez 2005 a,b, hereafter RM05a,b) and used to compute
the Li abundances (from equivalent widths taken from the literature). 
Because RM05a,b became public at a late stage in our refereeing process, we
did not update our input targets list, but instead we decided to discuss and
compare their and our results when relevant.

\section{Stellar parameters. I. Gravity, metallicity, and microturbulence}

Analysing lithium is not very difficult. Lithium appears in a stellar 
absorption spectrum with few transitions, namely the resonant line at 
670.7\,nm, and a much weaker signature at 610.4\,nm, only recently
explored in the most metal-poor stars (cf Bonifacio \& Molaro 1998,
Ford et al. 2002). The 670.7\,nm line falls in a clean spectral region,
especially in metal-deficient stars. From its equivalent width it is
easy to derive an abundance, once the stellar parameters of the object
under investigation have been determined. The sensitivity of the final
Li abundance to surface gravity, metallicity, and microturbulence is
not very significant (see below), whereas an uncertainty of $\pm$ 70\,K
in \teff\ (commonly quoted as a reasonable uncertainty on this parameter,
for solar-type stars) translates into $\pm$0.056\, dex on the
final lithium abundance. Because the effective temperature
is clearly the most critical parameter, it will be discussed separately, in
the next section, where we provide a more detailed description of what we
have learned from its derivation. Here, we will briefly comment on the other
input stellar parameters and how they were determined.

The {\it surface gravity} was first determined from an inspection of the (b-y)
{\it vs} c$_1$ diagramme (cf Fig.~2), which allowed us to assign a preliminary
\logg~value to each of our stars. These first-guess values were first checked
versus those quoted in the literature sources used to assemble our sample.
Then we finally attributed to each star the \logg~value deduced from its
position in the Herzsprung-Russel diagram (see \S~9.2).

\begin{figure}[t]
\rotatebox{0}{\resizebox{9cm}{!}{\includegraphics{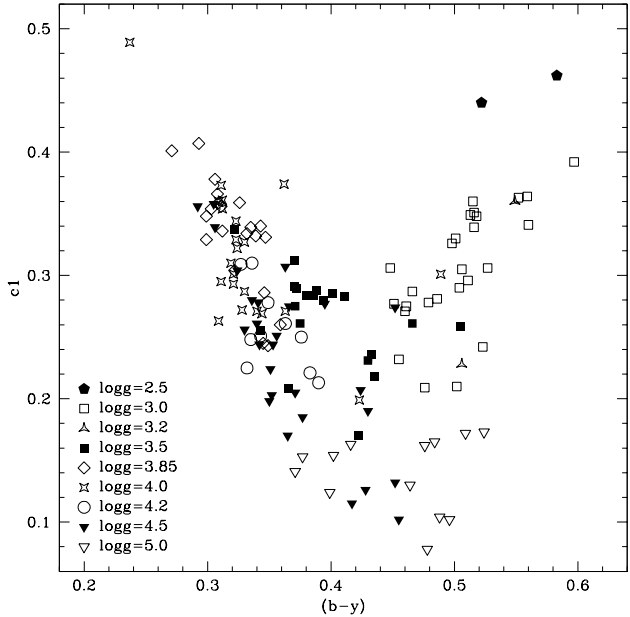}}}
\caption{The entire sample plotted in the (b-y) {\it vs} c1 diagram from
which we assigned the first set of gravities. Different symbols identify
different values of \logg, as summarised in the lower left corner of
the figure.}
\label{logg}
\end{figure}

In order to evaluate the sensitivity of the derived A(Li) abundances on the
stellar gravity, we performed our tests at \teff=5250~K and 6000~K, for
\logg=3.0 and 4.0, and for two metallicities, [Fe/H]=$-$1.0 and $-$2.5. On
the average, we found that a change of 1~dex in \logg\,affects the lithium
abundance by $\simeq$0.018~dex only, with very little dependence on the
effective temperature, or on the equivalent width of the lithium line. Our
findings agree very well with the common statements that the dependence of
A(Li) on the stellar gravity is negligible in the error budget (e.g.
Mel\'endez \& Ram\'{\i}rez 2004).

The {\it metallicity} was first selected from the same literature sources
from which we assembled our data sample. As one can see from inspecting
Table~3 the agreement between different literature sources (when available)
is in general quite satisfactory. Only in few cases, a large discrepancy
is present but was fortunately solved because observed spectra were in hand.
One such example is the star HIP\,88827, for which BM97 reported
[Fe/H]=$-$0.91 (taken from Alonso et al. 1996) and R96 measured $-$2.10:
from a high resolution, high S/N UVES spectrum taken at the VLT, its
metallicity has been recently derived from a reliable set of Fe\,II lines
and found to be $-$2.4 (Nissen et al. 2002). Because of the general good
agreement, our final metallicities are simply the weighted mean of all the
values available for a given star (with the exception, of course, of the very
few discrepant cases mentioned above). All the [Fe/H] values determined
spectroscopically from high resolution, high S/N spectra were double weighted.
The uncertainty on the final [Fe/H] values is the $\sigma_{n-1}$ of the
weighted mean (cf Table~6).

We cross-checked our metallicities also photometrically, via the calibration
of Schuster \& Nissen (1989, hereafter SN89 -- cf equation \#3), and taking
advantage of the availability of {\it ubvy}-$\beta$ photometry for the entire
sample. The satisfactory agreement we found was judged more than sufficient
for our test-purposes, therefore we did not explore any further the possible
systematic differences between metallicities derived spectroscopically and
photometrically. 

Although the Li abundance is only slightly dependent on the adopted
metallicity (0.2-0.3~dex uncertainties in [Fe/H] affect the Li abundance
less than 0.01~dex, cf R96), one has to remember that a possible error in
the metallicity may affect the A(Li) vs [Fe/H] trend, and more importantly
the \teff\, derived for that star and consequently its lithium abundance.
Our calculations confirm what already found by R99: if 
[Fe/H] is off by 0.15\,dex (a reasonable uncertainty for this parameter,
especially since we have assembled our sample from a variety of analyses),
the effect on \teff\, is almost negligible ($\approx \pm$20\,K) which implies
an uncertainty on the Li abundance of less than 0.02\,dex. This
sensitivity applies, of course, to the parameters space spanned by our
stars, with a tendency of finding larger dependences as the metallicity
increases. 

A similar, almost negligible dependence, is found also between Li and {\it
microturbulence}, for which $\pm$0.5\kms\, in $\xi$ correspond to
$\pm$0.005~dex in A(Li). Because of this negligible dependence and after some
checks of previous literature works, that included both dwarf and (sub)giant
stars, we decided to run all our calculations assuming $\xi$=1.5\kms. Because
of the very small dependence of A(Li) on microturbulence this choice gives
identical results to what was implemented by PSB93, who let $\xi$ varying
smoothly between 1.0 and 2.0 \kms\,going from the hotter to the cooler stars
of their sample. 

\section{Stellar parameters. II. The temperature scale and its weaknesses}

The main goal of any lithium analysis is to determine a fully consistent
temperature scale for all the targets under examination, as the lithium
abundance is strongly dependent on this stellar parameter. This approach
is usually considered a guarantee of the absence of spurious differences
possibly arising by having applied different criteria to the derivation
of the effective temperature. Ideally, one would like to determine
this parameter from first principles, i.e. to derive direct temperatures
for metal-poor dwarfs. In practice, this has been achieved sofar
for very few and very bright targets only (cf RM05a for a summary of what
is currently available).

\begin{table*}[ht]
\label{table4}
\caption{Photometry and reddening excesses. The whole table is available on-line} 
\scriptsize{
\parbox{220mm}{
\setlength{\tabcolsep}{0.15cm}
\begin{flushleft}
\begin{tabular}{rcccccrcccccccc}
\noalign{\smallskip}
\hline
\hline
\noalign{\smallskip}
 HIP    & (b-y) & c1 & m1 & $\beta$ & Ref$^{\mathrm{a}}$ & E(b-y)  & (B-V) & (B-V) & E(B-V) & E(b-y) & E(B-V) & E(b-y) & E(B-V) & E(B-V) \\ 
        &       &    &    &         &                    & $\beta$ & Hip   & Lit   & S98    & S98    & H97    & H97    & Lit    & BH    \\
\noalign{\smallskip}
\hline
\noalign{\smallskip}
   484 & 0.513 & 0.349 & 0.155 & \ld   & 2 &  \ld   & 0.787 & \ld   & 0.021 & 0.016 & \ld   & \ld   & \ld   & \ld   \\
   911 & 0.341 & 0.278 & 0.067 & 2.582 & 1 & -0.008 & 0.570 & 0.450 & 0.020 & 0.015 & 0.022 & 0.016 & 0.000 & 0.015 \\
  2413 & 0.549 & 0.360 & 0.078 & 2.730 & 2 &  0.167 & 0.792 & 0.793 & 0.395 & \ld   & 0.184 & 0.134 & \ld   & \ld   \\
  3026 & 0.332 & 0.334 & 0.087 & 2.598 & 1 & -0.002 & 0.465 & 0.460 & 0.036 & 0.026 & 0.034 & 0.025 & 0.000 & 0.015 \\
  3430 & 0.309 & 0.360 & 0.040 & \ld   & 2 &  \ld   & 0.401 & 0.390 & 0.723 & \ld   & 0.034 & 0.025 & 0.000 & \ld   \\
  \ld  & \ld   & \ld   & \ld   & \ld   & \ld & \ld  & \ld   & \ld   & \ld   & \ld   & \ld   & \ld   & \ld   & \ld   \\
\noalign{\smallskip}
\hline
\end{tabular}
\end{flushleft}
\begin{list}{}{}  
\item[$^{\mathrm{a}}$] {\it References}: 1. Schuster \& Nissen (1988); \\
1a. Schuster (2002) ({\it priv.comm.}); \\
2. Hauck \& Mermilliod (1998); \\
3. Laird, Carney, \& Latham (1988); \\
4. Ryan et al. (1999). \\
This legend refers to the table given in its entirety on-line.        
\end{list}{}{}
}
}
\end{table*}

Stellar temperatures can be determined spectroscopically (e.g. via profile
fitting of the wings of some of the Balmer lines, or from minimising the
slope between the iron abundance - as derived from Fe\,I lines - and their
excitation potential), or from photometry. Since we have assembled our data
sample from the literature (i.e. no newly observed spectra), photometry is
the only choice we have. Furthermore, because of the size of the sample only
Str\"omgren $uvby$-$\beta$ photometry (among the photometric indices most
sensitive to stellar temperatures) is available for all our stars, thanks
to the extensive photometry by Schuster \& Nissen (1988), supplemented
by unpublished photometry by Schuster ({\it private communication}) for
approximately 10\%\ of our stars. Table\,4 (available in its entirety only
on-line) summarises all the photometry we have used, together with the 
different colour excesses we have derived and that we will now discuss.
For reference and test purposes, it also includes (B-V) values
taken from Hipparcos (ESA 1997) and from the literature, but we remind the
reader that (B-V) is not a good temperature indicator.

We derived \teff\ from the $(b-y)_0$ colour index using the IRFM calibrations
of Alonso et al. (1996, 1999 plus the erratum from 2001) for dwarf and giant
stars (cf their equations \#6 and \#14 respectively). The evolutionary status
assigned to each object for the determination of the gravity (see
previous section) was used to decide if a star was a dwarf or a post-main 
sequence object.
In order to overcome the known problem of Alonso's calibration (i.e.
\teff\ diverging towards high values at the lowest metallicities, cf
R99, Nissen et al. 2002), we adopted a lower limit of [Fe/H]=$-$2.1 in the
equation. Two comments are mandatory here. Firstly, although it is
important to keep in mind this {\it a posteriori} solution when discussing
the findings on the most metal-poor stars of our sample, we note that Nissen
et al. (2002) found no worrysome behavior (due to the assumption of this lower
limit on [Fe/H]) when comparing effective temperatures derived via Alonso's
colour-\teff~calibrations from (b-y) and (V-K). Secondly, we note that RM05b
have argued for the first time that this characteristic of the Alonso's
calibration is not a problem, in the sense that it is not a numerical
artifact due to the quadratic dependence on [Fe/H] but it is intrinsic to
the IRFM. In their recent work, they see a similar effect not only in the
\teff\,{\it vs} (b-y) plane, but also for other colour indices. If confirmed,
this would imply that adopting a lower limit on the metallicity is not
justified, thus different effective temperatures would be derived. How
different can be seen from Figure 3, where we plot the effective
temperatures of all our stars with [Fe/H]$\leq-$2.1 derived from applying
or not this lower limit at $-$2.1. Knowing what the dependence of A(Li) on
\teff\,is, one can already have an idea of what the effect will be on our
final Li abundances. We will come back to this when discussing our results.

\begin{figure}[h]
\rotatebox{0}{\resizebox{9cm}{!}{\includegraphics{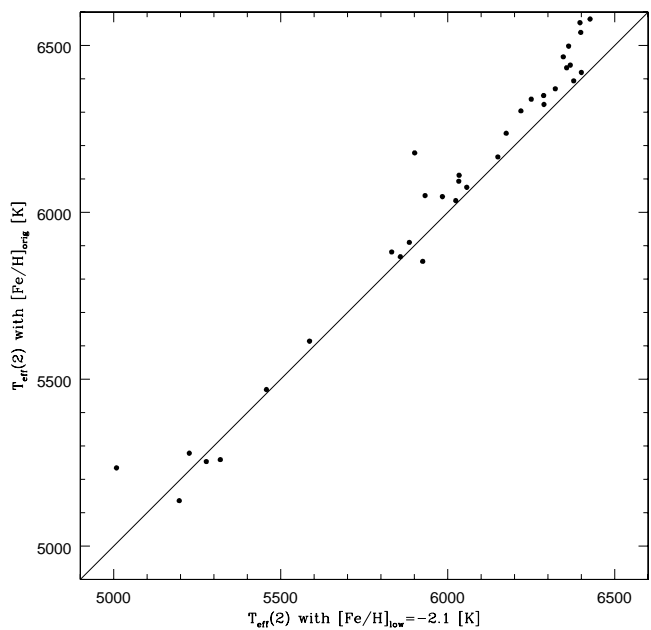}}}
\caption{A direct comparison of effective temperatures derived from the
Alonso's calibration, but adopting a lower limit of [Fe/H]=$-$2.1 for the
values plotted on the x-axis. Please note that \teff(2) refers to our final
set of effective temperatures (cf \S~6)}
\label{TfehM}
\end{figure}

The interstellar reddening excess was estimated from the $(b-y)_0-\beta$
calibration of SN89, including a zero-point correction of +0.005\,mag (Nissen
1994). The sample of stars from which Schuster \& Nissen (1988) derived the
abovementioned relation span the metallicity range $-2.5 \leq [M/H]
\leq +0.22$, and specific ranges of Str\"omgren colour indices, namely
$2.55 \leq \beta \leq 2.68$, $0.254 \leq (b-y) \leq 0.55$, $0.116 \leq
c1 \leq 0.540$, and $0.033 \leq m1 \leq 0.470$. 
  
However, despite (b-y), c1, m1 indices are available for all our stars,
some of them fall outside the ranges of validity for the calibrating
equations we have planned to use, and in few cases the $\beta$ index is
missing. For consistency, one is then left with two
possibilities: a) to reduce the sample to only those objects for
which the complete set of valid Str\"omgren photometric indices is
available (our sample would then be reduced to approximately 90 stars);
b) to replace the missing information with other methods and carefully
investigate the effects of such ``pollution'' (unavoidable in order
to keep the number statistics high) on the final output(s). Option a)
clearly represents the simplest path, and it will be used as our benchmark.
Here, instead, we want to describe in some detail the series of compromises
(i.e. sample pollutions) we had to introduce in order to keep working with
our whole data sample. At the end, we will compare a) to b), and discuss
how these two approaches affect the final Li abundances.

\subsection{The first pollution: the reddening excess without the beta index. 
Definition of the $\beta$-sample} 

Estimating the reddening correction is probably the weakest point of any
photometrically-based \teff\ scale. Interstellar reddening excesses are
rarely quoted and discussed in spectroscopic analyses, despite they can
affect significantly the determination of any stellar parameter, the
effective temperatures in particular. In order to be consistent within
the Str\"omgren photometry framework, we decided to use the $(b-y)_0-\beta$
relation to evaluate the interstellar reddening excess. However, for 24
stars ($\sim$20\% of the sample), the $\beta$ index was not found. A
common solution shared by several analyses has been to assume zero
reddening, based on the fact that the stars likely belong to the solar
neighborhood. Alternatively, one could derive E(b-y) from other colour
excesses, if available. Either way, this is an approximation, that we
consider as the first compromise on our data-sample. When referring to
this sub-sample of stars, we will call it the 
{\sl $\beta$-sample}. 

We decided to derive E(b-y) from E(B-V) via the formula $E(B-V) = 1.35\times
E(b-y)$ (Crawford 1975), which is based on a 1/$\lambda$ reddening law and
on the central wavelengths of the bandpasses. For the E(B-V) colour
excesses we simply averaged the E(B-V) values derived from IR Dust Maps
(Schlegel et al. 1998, hereafter S98) and the models of large scale visual
interstellar extinction by Hakkila et al. (1997, hereafter H97). If these
two sources (which will be extensively discussed in \S~5.4) were found to
diverge significantly (e.g. in the case of HIP 49616, for which we find 0.177
from S98 maps and 0.024 from H97 models), that object was dropped from our
list. The E(B-V)$_{Lit}$ values were given a much lower weight, being their
original source not always available. However, when found in agreement with
the other two E(B-V) sources, they were included in the straight average.
Because this solution includes a mixture of E(B-V) sources, from now on we
will refer to these values as E(B-V)$_{mix}$. Out of the 24 stars we have
without the $\beta$ index, 14 were rescued. Figure~4 shows a direct
comparison between (b-y)$_0$ values which have been corrected for reddening
derived respectively from $\beta$ (on the x-axis) and from E(B-V)$_{mix}$
values and Crawford's formula (on the y-axis). This plot includes all the
stars of our sample for which both methods could be safely applied. There
is clearly some scatter around the 1:1 relation (on the order of 0.015mag),
and a systematic tendency of deriving larger reddenings from the
indirect formula.   

\begin{figure}[h]
\rotatebox{0}{\resizebox{9cm}{!}{\includegraphics{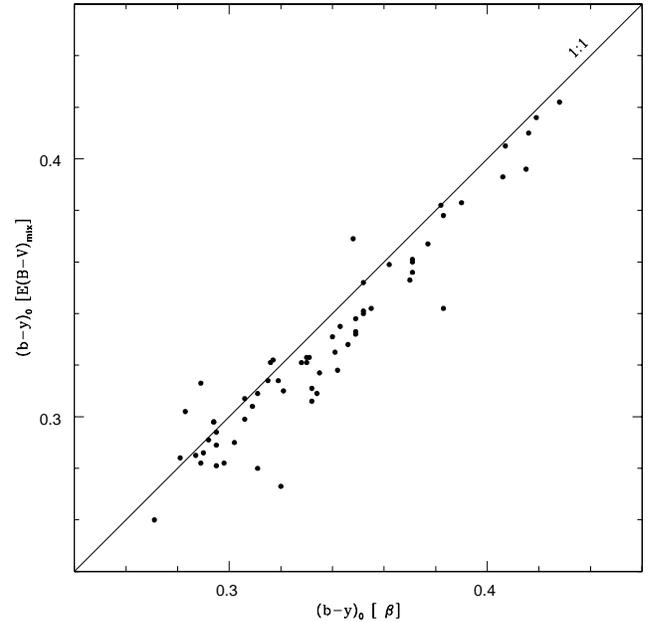}}}
\caption{A direct comparison of the (b-y)$_0$ index, de-reddened respectively
from the $\beta$-index (x-axis) and from the E(B-V)$_{mix}$ values averaged
from several sources (y-axis; see text for details), via Crawford's (1975)
formula}
\label{by0}
\end{figure}

\subsection{The second pollution: the validity of the reddening $(b-y)_0-
\beta$ relation. Definition of the {\sl ubvy}-sample}

We mentioned above that the equation to be used in the derivation of the
interstellar reddening is valid (i.e. has been tested) only for specific
ranges of the Str\"omgren indices. Taking the photometry at face value,
our sample includes a total of 35 stars, for which at least one of the
Str\"omgren indices does not fulfill these criteria: 7 stars have the
(b-y) index outside the 0.254--0.550 interval, 5 stars have the c1 index
outside 0.116--0.540, another 8 have the m1 index falling outside
0.033--0.470, and 21 have the $\beta$\,index outside 2.55--2.68 (six of
which also have some of the other indices off). 

For the 15 (i.e., 21$-$6) stars with $\beta$ (only) outside the allowed
range, we were able to apply the same solution described in \S~5.1, i.e.
we derived E(b-y) from E(B-V)$_{mix}$ values, to 6 of them. If the remaining 9
stars are dropped, together with those 20 ($7+5+8$) that have (b-y) or m1 or
c1 outside the allowed ranges, we are then left with a total of 111 stars, of
which 20 are ``polluted'' because their E(b-y) was derived from E(B-V)$_{mix}$
(of the initial 39=24+15 stars that belonged to this sample, 5 were dropped
because one or more of their Str\"omgren indices fell outside the allowed
ranges). We note, however, that so far we took all the photometric indices
at face value, whilst each of them has its own associated uncertainty. If
one were to take this into account, then the application of the validity
ranges would allow some flexibility. For some stars (7 out of 20) such
approach seems very reasonable: HIP103337, for instance, has the (b-y) index
off by 0.002~mag, which translates to a 10~K effect; all `m1' drop-outs
(except two) have their photometric index off by only few thousandths of a
magnitude (at most by 0.006~mag), which affect their final effective
temperatures between 3.5~K and 20~K. Because of these considerations, we
decided to keep these 7 objects in a separate (polluted) sample, which we
call the {\sl {\it ubvy}-sample}. In summary, we are then left with a total of 118
stars. 

Should we have considered also the $\beta$ values with a 3-digits precision,
six more targets would now belong to the {\it ubvy}-sample. Since this could
influence our final discussion of the A(Li) plateau, we will take a closer
look at them once their Li abundances have been derived (cf \S~7).

\subsection{The third pollution: How and when to apply the reddening
corrections. Definition of the four sets of T$_{\rm eff}$} 

Knude (1979) showed that interstellar reddening is caused primarily by
small dust clouds with a typical reddening of $E(b-y)\simeq 0.03$. If
true, this would make any correction for reddening values smaller than
0.03 almost meaningless. This is why in the past it has been common
practice to correct only those stars for which the derived excess was
comparable to or larger than this value. SN89, for
instance, chose $E(b-y)=0.025$ as their reference value and performed
the corrections $(b-y)_0 = (b-y) - E(b-y)$, $c_0 = c_1 - 0.2E(b-y)$, and
$m_0 = m_1 + 0.3E(b-y)$ for all stars with $E(b-y) \geq 0.025$. 

However, the commonly accepted picture of the nature and appearence of
the interstellar reddening has significantly evolved since Knude's work
and seems to suggest a patchy distribution of interstellar dust (implying
that values smaller than E(b-y)=0.03 may be real) with a void of about
70-75\ pc around the Sun (e.g. Lallement et al. 2003). In principle, one
should then apply the interstellar reddening excesses derived from the
$(b-y)_0$ relation to all the stars. In practice, one has to face the
problem of correcting also those stars for which negative E(b-y) values
are derived (28 objects in our case). This means switching to the $\beta$
index as the main \teff\ indicator, thus losing precision for those stars
that are close enough to be in the void around the Sun (Nissen, {\it priv.
comm.}). This explains why it is still common procedure to choose a minimum
E(b-y) threshold below which the colour indices are not corrected for
reddening excesses. For instance, Nissen et al. (2002) have arbitrarily
chosen E(b-y)=0.015, which corresponds to twice the sigma of E(b-y). 

Being unsure of what is the best approach to follow, we tested the effects
of the abovementioned solutions by deriving different sets of
temperatures, under the following assumptions: 1. all stars were de-reddened
[\teff(1)]; 2. all stars were dereddened except those with negative E(b-y)
values [\teff(2)]; 3. all stars were dereddened except those with E(b-y)
$\leq$ 0.01 [\teff(3)]; 4. all stars were dereddened except those with
negative E(b-y) values and those with E(b-y) $\leq$ 0.01 and d$<$70~pc
[\teff(4)]. Our four derived sets of effective temperatures are compared
in Fig.~5 (\teff\,(1) always on the x-axis) and show very little differences,
the smallest in the case of \teff(1) compared to \teff(2) ($\pm$27~K around
the mean, cf top panel). Looking at these two sets of \teff\, in
more detail (the crosses in the top panel of Fig.~5 identify the 28 differing
stars), one notices that for one third of these stars there is practically
no difference (these are the stars for which very small reddening excesses,
in absolute value, were found to be negative). Except for 3 targets, for
which the difference in effective temperature is larger than 100~K, all
the others are within $\pm$50~K, with a systematic tendency of the \teff\,(2)
values to be slightly higher than \teff\,(1) ones.

\begin{figure}[t]
\includegraphics[width=9cm,height=12cm]{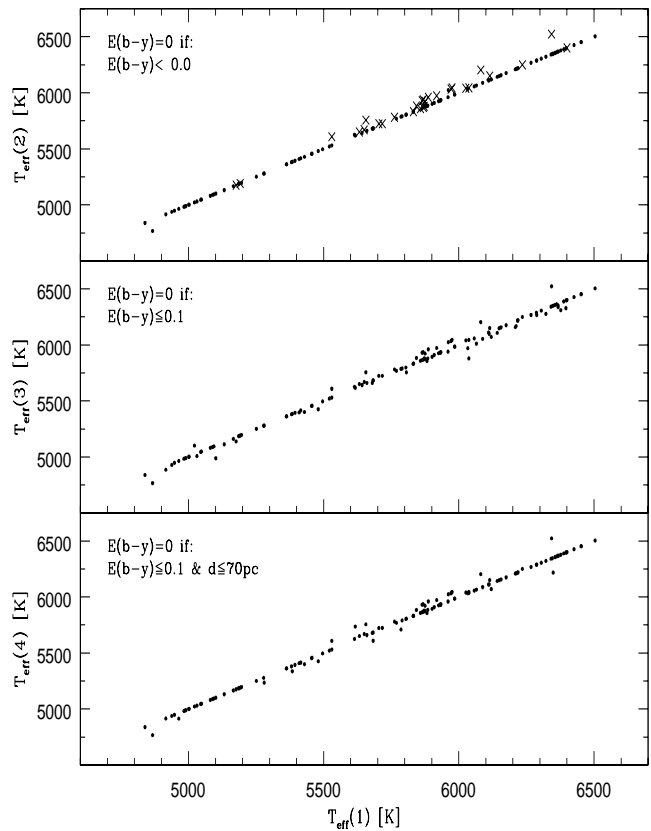}
\caption{How the sets of temperatures \teff(2), \teff(3), and \teff(4) (on
the y-axis; see text for more explanations) compare to \teff(1) (on
the x-axis). Crosses in the {\it top} panel represent the 28 objects for
which the two criteria (\#1 and\#2) give different values}
\label{Tred}
\end{figure}
 
\subsection{More thoughts on the interstellar reddening excess} 

The fact that a very small difference in E(b-y) (such as 0.005\,mag,
which accounts approximately for half of the common uncertainty) translates
already into 35\,K in effective temperature is a strong indication that
accounting for interstellar reddening excesses plays an important role in
deriving an accurate photometrically-based \teff\ scale - especially when
the abundance of the element(s) under investigation is very sensitive to
\teff\,like in the case of lithium. Therefore, it is important to comment
also on the other sources of interstellar reddenings available in the
literature. 

In order to complete our comparison tests, we decided to derive
reddening values also from the most recent tabulated values, i.e. those
derived from infrared mapping of the dust emission distribution (Schlegel
et al. 1998, S98 for short), and from the models of large scale visual
interstellar extinction (Hakkila et al. 1997, hereafter H97).
This was done not only to thoroughly test our photometrically-based reddening
values, but especially to evaluate a posteriori the effect of mixing
different sources of reddening on the derived temperature scale (a common
approach to many spectroscopic abundance analyses). 

Schlegel et al. (1998) estimated the dust column densities from the COBE/DIRBE
and IRAS/SISSA infrared maps of dust emission over the entire sky, and
transformed them to reddenings by using colours of elliptical galaxies. In
other words, these maps give reddening values as if the objects lie outside
the Galaxy, hence they may overestimate the real reddening, especially for
relatively nearby objects. Also, at low latitudes ($|b|<5\deg$) the removal
of IR point sources is not optimal, hence the derived reddening values may
be strongly affected. The quoted errors are of the order of 16\%. 

Hakkila et al. (1997) instead, developed a numerical algorithm to model the
large scale Galactic clumpy distribution of obscured interstellar gas and
dust by using published results of large-scale visual interstellar extinction.
It is concentrated towards the Galactic plane and it varies as a function of
Galactic longitude and latitude. These estimates depend on the assumed
distance, which is one of the input parameters to the algorithm. A word of
caution concerns its inability in identifying small scale (less than 1$\deg$)
extinction variations, and the fact that reddening estimates for mid-Galactic
latitudes ($7\deg \leq b \leq 16\deg$) and for distances between 1 and 5 kpc
in the Galactic plane are more unsecure. The quoted errors are not very
meaningful since they represent the mean of the errors as reported in the
original studies, hence they are likely overestimated. 

Table\,4 presents an overview on how reddening excesses derived from different
methods compare to each other. Columns 2 to 7 report Str\"omgren photometry
and E(b-y)$_{\beta}$ values as derived from the (b-y)$_0$-$\beta$ calibration
(see previous sub-sections), while columns 8 and 9 list the Hipparcos-
and literature-based (B-V) values. Column 10 reports the E(B-V) values as
derived from the S98 maps, while column 12 lists the values as derived from
the H97 algorithm. The corresponding E(b-y), derived via the Crawford (1975)
relation are reported in columns 11 and 13 respectively.

Two remarks are important. First, some very high values of E(B-V) are derived
from the S98 maps (by using the IDL code made available by the same authors),
which look unrealistic when compared to all the other E(B-V) values. Since
the main purpose of our tests is to have some feeling on the possible scatter
introduced by mixing reddening values taken from different sources, without
being biased by outliers, we did not investigate these high values any
further.
Hence, they have been discarded from all comparison figures and tests.
However, one can easily identify them in Table\,4, since no corresponding
E(b-y) value was derived (same applies also to Hakkila-based E(B-V) values -
though for significantly fewer stars). One should also note that in Table\,4
there remain some suspiciously high E(B-V) values. 

\begin{figure}[t]
\rotatebox{0}{\resizebox{9cm}{!}{\includegraphics{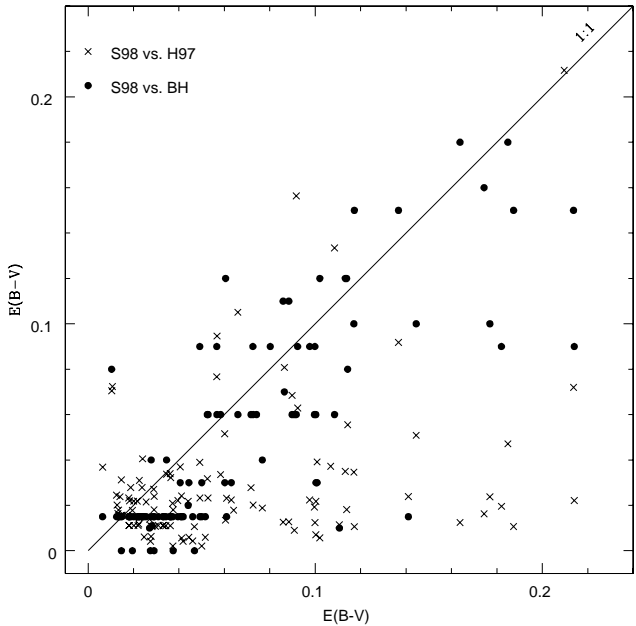}}}
\caption{Comparison between E(B-V) values derived from different methods: {\it
filled dots} show how E(B-V) colour excess compare when derived from the
InfraRed Dust Maps of Schlegel et al. (1998, on the x-axis) and of
Burstein \& Heiles (1982, on the y-axis); {\it crosses} represent a similar
comparison between E(B-V)$_{S98}$ and E(B-V)$_{H97}$ (on the x- and
y-axis respectively).}
\label{EBVcomp}
\end{figure}

\begin{figure}[t]
\rotatebox{0}{\resizebox{9cm}{!}{\includegraphics{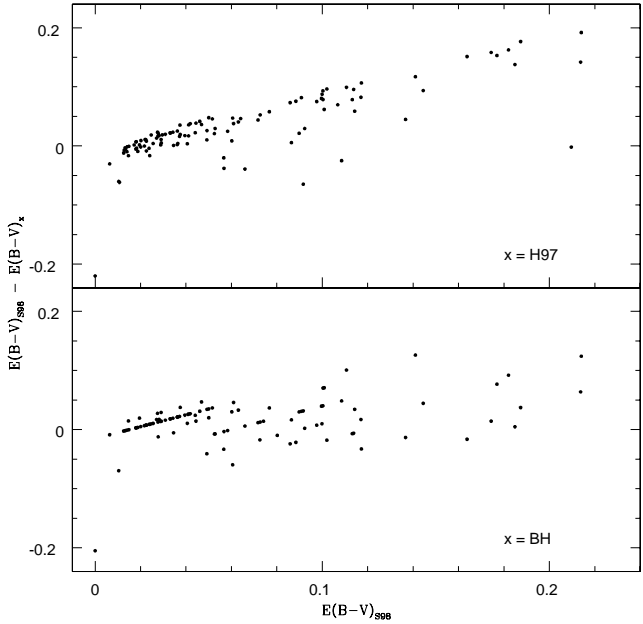}}}
\caption{Comparison between E(B-V)$_{S98}$ (on the x-axis) and differences
between E(B-V) values derived from various sources (identified from the
acronym given in the right bottom corner of each pan), always using S98 as
the reference}
\label{EBVcompdiff}
\end{figure}

Secondly, in order to survey as many choices of reddening as possible,
Table\,4 includes also E(B-V) values as found in the literature sources from
which we assembled our data sample 
(column 14 labeled E(B-V)$_{Lit}$) and as
derived from the neutral hydrogen {\ion{H}{i}}  column density distribution
of Burstein \& Heiles (1982, BH for short -- 
column 15 labeled E(B-V)$_{BH}$)
in correlation with deep galaxy counts. A partial summary of Table\,4 is
provided in Fig.~6, where two comparisons are plotted simultaneously: the
filled circles show the relation between E(B-V) values derived from the S98
(on the x-axis) and from the BH maps (on the y-axis). The crosses represent
the comparison between E(B-V) values derived from the S98 IR dust maps (on
the x-axis) and from the H97 model (on the y-axis). The 1:1 relation is
plotted for comparison. From this figure, one immediately notices the lack
of a tight correlation between the E(B-V) values derived from the maps of
Schlegel et al. and Burstein \& Heiles. The comparison with H97 clearly
shows that Hakkila's model of the Galactic interstellar dust distribution
tends to give reddening values much lower than the S98 IRDM. Another way to
look at these comparisons is by inspecting Fig.~7, where the difference
between the different reddening estimates (always with respect to S98 values)
are plotted versus E(B-V)$_{S98}$. From this figure it becomes clear that
there are systematic differences between S98 reddening values and the other
two estimates, although there does not seem to be any dependence on the
distance, except for a slightly larger dispersion at small distances (cf
Fig.~8). 

\begin{figure}[t]
\rotatebox{0}{\resizebox{9cm}{!}{\includegraphics{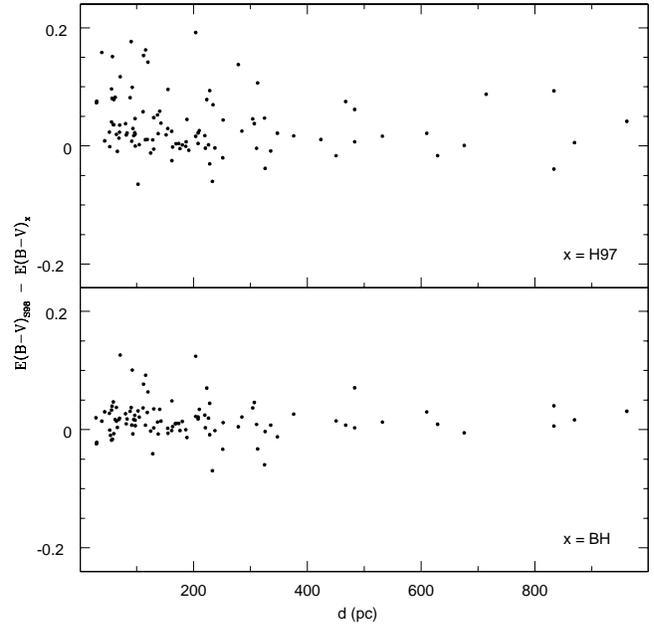}}}
\caption{Comparison between E(B-V) values derived from various sources
(identified from the acronym given in the right bottom corner of each pan)
and the distance of the object, as derived from Hipparcos parallaxes.}
\label{EBVcompdiffd}
\end{figure}

In summary, based on the abovementioned arguments, it is very hard to defend
the position that we know reddening better than 0.007-0.010mag (2$\sigma$)
which correspond already to affecting the temperature determination by
50-70\,K (and in turn the lithium abundance by $\sim$0.05~dex).

\section{Our final choice: temperature and data-sample(s)}
Because lithium abundances are mostly sensitive to the choice of the stellar
temperature, our main goal has so far focused on how to derive a temperature
scale as consistent as possible. In order to do that, we chose to derive
photometric temperatures since our sample has been assembled from different
literature sources. Figure~9 shows how well our final set of \teff\,correlates
with (b-y)$_0$ (i.e., the Str\"omgren index our temperature scale is based on).
During this process (cf \S~5) we faced some of the major drawbacks of
such determinations, namely reddening and applicability of colour-\teff\,
relations. These are summarised in Figs.~3-8. By inspecting Fig.~5 and
noticing how small the differences between \teff(1) and \teff(2) are,
we have selected \teff(2) as our final set of effective temperatures
(listed in the second column of Table~5, labeled \#2). As a reminder, this
was derived by de-reddening only those stars for which E(b-y) was found to
be positive. 

\begin{figure}[t]
\rotatebox{0}{\resizebox{9cm}{!}{\includegraphics{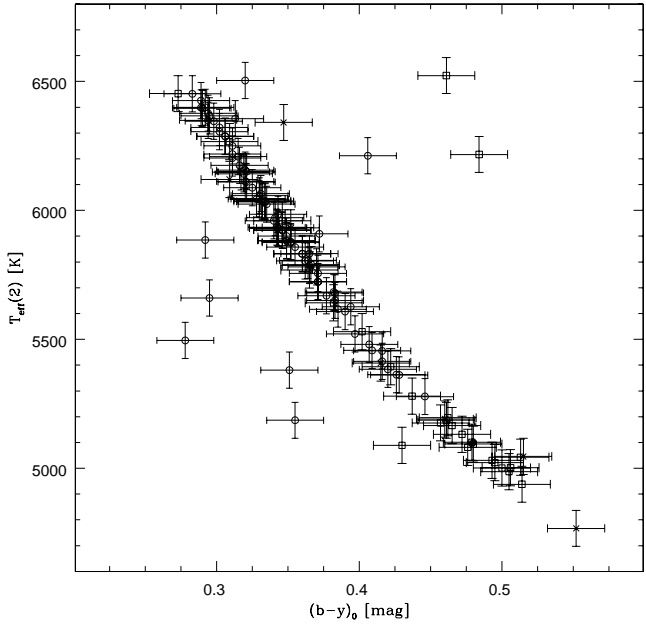}}}
\caption{How our final set of temperatures \teff(2) correlates with the
de-reddened Str\"omgren index $(b-y)_0$, demonstrating the strong sensitivity
of the index to effective temperature.}
\label{teffby0}
\end{figure}

\begin{table}[t]
\label{table5}
\footnotesize{
\caption{Final temperatures as derived from different reddening corrections.
The whole table is available on-line}
\begin{flushleft}
\begin{tabular}{rcccccc}
\hline
\hline
\noalign{\smallskip}
 HIP   & \multicolumn{6}{c}{Effective Temperature Scales (K)} \\
       & 2 & 1 & 3 & 4 & S98 & H97 \\
\hline \\[0.02mm]
\multicolumn{7}{c}{Sample \#1: the {\sl clean} sample}\\ 
\noalign{\smallskip} 
\hline
\noalign{\smallskip}
   911 & 5972 & 5918 & 5972 & 5972 & 6076 & 6065 \\
  3026 & 6040 & 6026 & 6040 & 6040 & 6221 & 6214 \\
  3446 & 5901 & 5901 & 5894 & 5901 & 5991 & 5995 \\
  3564 & 5683 & 5683 & 5683 & 5683 & 5974 & 5679 \\
  8572 & 6287 & 6287 & 6287 & 6287 & 6265 & 6287 \\
  \ld  & \ld        & \ld  & \ld  & \ld  & \ld  & \ld  \\
\noalign{\smallskip}
\hline \\ [0.02mm]
\multicolumn{7}{c}{Sample \#2: the {\sl $\beta$} sample} \\ 
\noalign{\smallskip}
\hline
\noalign{\smallskip}
   484 & 5064 & 5064 & 5064 & 5064 & 5064 & \ld  \\
  3554 & 5008 & 5008 & 5008 & 5008 & 5040 & 5022 \\
  4343 & 5064 & 5064 & 5064 & 5064 & 5069 & 5064 \\
  8314 & 6430 & 6430 & 6430 & 6430 & \ld  & 6444 \\
 13749 & 4965 & 4965 & 4965 & 4965 & 4996 & 4965 \\
  \ld  & \ld  & \ld  & \ld  & \ld  & \ld  & \ld  \\
\noalign{\smallskip}
\hline \\ [0.02mm]
\multicolumn{7}{c}{Sample \#3: the {\sl ubvy} sample} \\ 
\noalign{\smallskip}
\hline
\noalign{\smallskip}
 12807 & 5932 & 5932 & 5932 & 5932 & 6872:& 5959 \\
 83320 & 5984 & 5984 & 5984 & 5984 & 6251 & 5861 \\
 87062 & 5909 & 5909 & 5909 & 5909 & \ld  & 5905 \\
 91129 & 6217 & 6217 & 6217 & 6217 & 7081:& 6237 \\
103337 & 4885 & 4885 & 4885 & 4885 & 4876 & 4876 \\
  \ld  & \ld  & \ld  & \ld  & \ld  & \ld  & \ld  \\
\noalign{\smallskip}
\hline
\end{tabular}
\end{flushleft}
}
\end{table}

Table\,5\footnote{only available in its entirety on-line} summarises
all sets of temperatures that have resulted from our several
digressions: columns
2 to 5 report the effective temperatures derived by applying the reddening
$(b-y)_0-\beta$ relation and making different assumptions about it (cf
\S~5.3), whereas columns 6 and 7 represent sets of temperatures as
derived by using different sources of reddening. The latter two columns
are useful comparisons to check how much scatter and slope on the
A(Li)-plateau could originate just by mixing different reddening sources
in the same analysis. Figure~10 clearly shows how different the derived
temperatures can be when the reddenings are derived from different methods:
if they are taken from the IRDM of S98, for instance, the derived temperatures
are significantly higher ($+$178~K on the average, with a dispersion around
the mean of $\pm$236~K). Of course, depending on for which targets one may
need to select a different source of reddening excess, there may well be some
artificial scatter emerging among their Li abundances. 

Another relevant comparison to make would be the one between our
\teff\,scale and the \teff\,values used in the original literature sources
from which our list of targets was assembled. We present these comparisons
in Fig.~11, in the form of \teff(2) $-$ \teff(X) vs \teff(2), where \teff(2)
is our preferred and finally selected set of temperatures and \teff(X)
represents the temperature scales used in the original works from which our
list of targets was assembled (cf Table~2). Here, we selected to plot those
literature analyses from which we took the largest numbers of stars, except
for the bottom panel where the comparison with the very recent RM05a
\teff\,scale is presented. We remind the reader that our list of targets
does not include objects from this work because the analyses were carried
out almost simultaneously. As one can see, our temperature scale (\teff(2))
tend to be always higher, the largest difference being with PSB93 (on average
$+$158~K, with a dispersion around the mean of $\pm$136~K). The smallest
differences, on average, are between us and BM97 ($+$6~K) and RM05a
(practically zero), but the dispersions around these means remain on the
order of $\pm$100-150K.

\begin{figure}[t]
\rotatebox{0}{\resizebox{9cm}{!}{\includegraphics{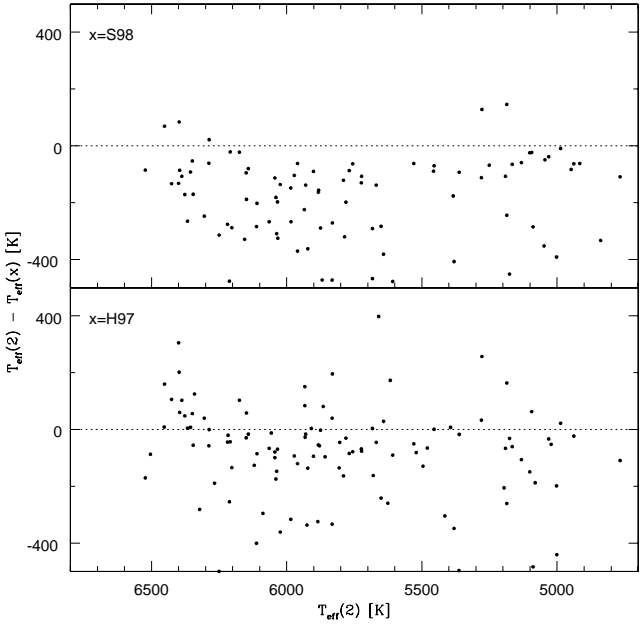}}}
\caption{Comparison between our final set of effective temperatures \teff(2)\
(on the x-axis) and the temperatures computed by using E(b-y) as derived from
E(B-V)$_{S98}$ and E(B-V)$_{H97}$ (on the y-axis, from top to bottom)}
\label{LiTeff}
\end{figure}

\begin{figure}[h]
\rotatebox{0}{\resizebox{9cm}{!}{\includegraphics{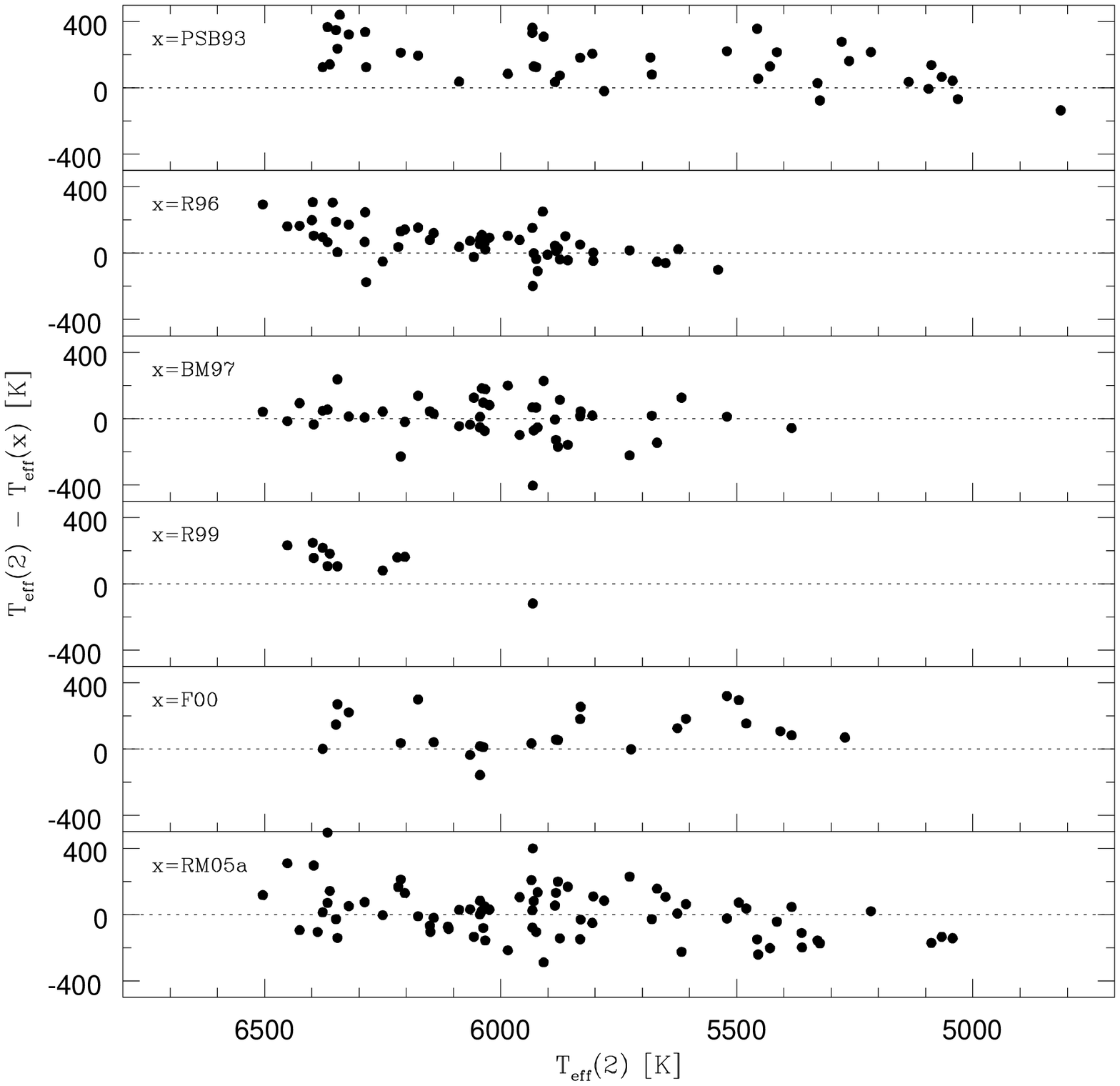}}}
\caption{Comparison between our final set of effective temperatures \teff(2)\
(on the x-axis) and the temperature scales used/derived in the original literature
sources (on the y-axis). Acromyms in the upper left corner identify which analysis
is used for the comparison.}
\label{LiTeff}
\end{figure}

Also, in Table\,5 all our targets are grouped in three
different sub-samples. As stated at the beginning of \S~5, in order to be
fully consistent with the analytical method one chooses to follow, one is
usually forced to work with a much smaller sample of stars compared to the
initial data-set: in our case, 91 stars compared to the original 146. In
order to avoid this drastic reduction, we explored alternative solutions
which allowed us to retain a larger number of objects 
(118), but at the price of contaminating part of the sample as follows:\\

{\it Sample \#1} is the {\sl clean} sample: it includes 91 stars, for which
the complete set of Str\"omgren photometric indices are available, and for
which the SN89 calibrations can be successfully applied.

{\it Sample \#2} is the {\sl $\beta$} sample: it includes 20 stars, for
which the reddening value E(b-y) was derived from averaging different
sources of E(B-V) values (S98, H97, and literature - cf \S~5.1 for details
on how this average was performed), via Crawford's formula (1975). We
note that no correction was applied to these stars to compensate for the
offset seen in Fig.~4.

{\it Sample \#3} is the {\sl ubvy} sample: it includes 7 stars, for which one
of the {\sl ubvy} photometric indices (b-y, c1, m1) falls just slightly
outside (see last paragraph of \S~5.2) the allowed intervals for the
application of the Nissen \& Schuster (1989) calibrations.

\begin{table*}[t]
\label{table6}
\caption{Final Li abundances as derived by using different sets of \teff. 
The whole table is available on-line} 
\scriptsize{
\parbox{220mm}{
\begin{flushleft}
\begin{tabular}{rcccccccccccc}
\hline
\hline
HIP & bin & [Fe/H] & $\sigma$(Fe) & logg & EW & $\pm1\sigma$ & A(Li)$_{LTE}$ & A(Li)$_{NLTE}$ & $+1\sigma$ & $-1\sigma$ & A(Li)$_{S98}$ & A(Li)$_{H97}$ \\
    &     & dex    & dex          & \cms & \mA & \mA         & dex           & dex            & dex        & dex        & dex           & dex \\
\hline \\
\multicolumn{13}{c}{Sample \#1: the {\sl clean} sample} \\ 
\noalign{\smallskip} 
\hline
\noalign{\smallskip}                          
   911 & & -1.84 & 0.15 & 4.50 & 30.8   & 3.5 & 2.117    & 2.112    & 0.042 & 0.067 & 2.200    & 2.191 \\
  3026 & & -1.25 & 0.07 & 3.85 & 45.0   & 6.0 & 2.428    & 2.418    & 0.078 & 0.086 & 2.573    & 2.567 \\
  3446 & & -3.50 & 0.10 & 4.50 & 27.0   & 3.9 & 1.973    & 1.986    & 0.074 & 0.076 & 2.045    & 2.048 \\
  3564 & & -1.27 & 0.15 & 3.50 & 35.2   & 3.5 & 2.011    & 2.054    & 0.056 & 0.076 & 2.244    & 2.008 \\
  8572 & & -2.51 & 0.01 & 3.85 & 27.0   & 1.4 & 2.257    & 2.236    & 0.026 & 0.027 & 2.239    & 2.257 \\
  \ld  & & \ld   & \ld  & \ld  & \ld    & \ld & \ld      & \ld      & \ld   & \ld   & \ld      & \ld   \\
\noalign{\smallskip}
\hline \\
\multicolumn{13}{c}{Sample \#2: the $\beta$ sample} \\ 
\noalign{\smallskip}
\hline
\noalign{\smallskip}
   484 &   & -1.23 & 0.15 & 3.00 & 12.1   & 3.5 & 0.953    & 1.101    & \ld   & \ld   & 0.949    & \ld   \\ 
  3554 &   & -2.87 & 0.15 & 3.00 & 17.7   & 3.5 & 1.018    & 1.153    & 0.087 & 0.104 & 1.044    & 1.238 \\ 
  4343 &   & -2.08 & 0.15 & 3.00 &  9.1   & 3.5 & 0.766    & 0.912    & 0.150 & 0.219 & 0.878    & 0.680 \\ 
  8314 & ? & -1.68 & 0.09 & 4.00 & 27.0   & 3.0 & 2.378    & 2.337    & 0.053 & 0.058 & 2.262    & 2.523 \\ 
 13749 &   & -1.62 & 0.15 & 3.00 & 14.6   & 3.5 & 0.876    & 1.039    & 0.102 & 0.128 & 0.901    & 0.870 \\ 
  \ld  &   & \ld   & \ld  & \ld  & \ld    & \ld & \ld      & \ld      & \ld   & \ld   & \ld      & \ld   \\
\noalign{\smallskip}
\hline \\
\multicolumn{13}{c}{Sample \#3: the {\sl ubvy} sample} \\ 
\noalign{\smallskip}
\hline
\noalign{\smallskip}
 12807 &   & -2.87 & 0.22 & 4.50 & 22.9   & 3.0 & 1.918    & 1.929 & 0.066 & 0.068 & 2.676    & 1.940 \\
 83320 &   & -2.56 & 0.15 & 3.50 & $<$5.0 & \ld & $<$1.265 & $<$1.287 & \ld & \ld  & $<$1.603 & 1.167 \\
 87062 &   & -1.67 & 0.23 & 4.50 & 31.5   & 4.0 & 2.109    & 2.111 & 0.088 & 0.069 & 2.109    & 2.106 \\
 91129 & * & -2.96 & 0.10 & 4.50 & 27.3   & 2.5 & 2.208    & 2.191 & 0.045 & 0.054 & 2.899    & 2.224 \\
103337 &   & -2.07 & 0.15 & 3.00 & 25.8   & 3.5 & 1.025    & 1.197 & 0.069 & 0.072 & 1.179    & 0.877 \\  
  \ld  &   & \ld   & \ld  & \ld  & \ld    & \ld & \ld      & \ld   & \ld   & \ld   & \ld      & \ld   \\
\noalign{\smallskip}
\hline
\end{tabular}
\end{flushleft}
\begin{list}{}{}
\item [?] identifies a suspected binary (Latham et al. 2002, Carney et al.
1994, 2003). 
\item[*] identifies a confirmed single- or double-lined binary (from
Latham et al. 2002, Carney et al. 1994, 2003).  \\
Note that the complete version of the table (available only on-line) reports
the complete legend of symbols and references
\end{list}{}{}
}
}
\end{table*}

\section{The lithium abundance}
The final assessment of how relevant and important all our 
tests have been can be made only after comparing the lithium abundances
derived from the different sets of temperatures and for the different
sub-samples. 

The lithium abundance for all the stars of our sample was determined from
the equivalent widths (EWs) of the 670.7\,nm line as reported in the
literature works from which we assembled the data sample. Table\,6\footnote{available
in its entirety on-line} lists the mean equivalent width and its
1$\sigma$ uncertainty that were used in our computations of the Li abundance:
we opted for the mean value because of a satisfactory overall agreement found
in the literature (cf Table\,3, for the corresponding references, listed in
the last column of the table). 

The lithium abundance was derived under the assumption of Local Thermodynamic
Equilibrium  (LTE) using Kurucz (1993) WIDTH9 and model atmospheres with the
overshooting option switched off (cf Castelli et al. 1997 for the models,
and Molaro et al. 1995 for comparisons between different versions of Kurucz
model atmospheres). The gf-value we used for the Li~I is 0.171, for which the
VALD database reports an accuracy of 3\%. 
 
As Carlsson et al. (1994) have shown, the LTE approximation when deriving Li
abundances for cool stars is not a realistic representation of the physics
present in the atmospheric layers where the 670.7~nm line forms. Since
non-LTE corrections vary in sign and size when spanning a large range of
stellar parameters (being larger for cooler stars), ignoring these
corrections may clearly affect any interpretation of the A(Li)-plateau and
its possible slope with effective temperature and metallicity. Therefore,
NLTE corrections were computed with the interpolation code made available
by Carlsson et al. (1994) and applied to our LTE A(Li) abundances. We note
that no NLTE correction could be derived for those stars with A(Li)$_{LTE}$
abundances smaller than 0.6, because the interpolation code works on a
given range of input parameters (for instance, in the case of the Li
abundance, the range is A(Li)=0.6--4.2). Furthermore, despite these ranges
of input parameters, not all the combinations are covered in the table which
contains the tabulated coefficients from which the NLTE corrections are
computed. In our sample, we had only few of these cases, namely two stars
had a metallicity lower than the minimum threshold ($-$3.0), and another
couple of stars had their effective temperatures slightly higher of the
maximum threshold (by 4 and 23~K respectively). For these objects, we
rounded off their parameters to the nearest allowed value, and computed
the NLTE correction. 

The resulting LTE Li abundances (both LTE and NLTE are given only for our
final set of \teff) are presented in Table\,6, where we give also other
relevant parameters like the metallicity and its 1$\sigma$ uncertainty as
derived from a critical analysis of the literature. 
Three different sets of lithium abundances are reported for each target,
depending on which set of \teff\, was used (cf Table\,5). For the A(Li)
values (which were derived from our final set of temperatures, i.e. \teff(2))
we also give the associated $\pm$1$\sigma(EW)$ error 
(cols. 9, 10). Similar uncertainties apply also to the lithium abundances 
listed under the last two columns of the table.  

\subsection{What is the best achievable accuracy ?}
The accuracy of any abundance determination mainly depends on the
following factors: the quality of the observational sample (for the
continuum placement and the measurement of equivalent widths), the choice
of the stellar parameters characterising each star of the analysed sample,
the atomic physics (e.g. the oscillator strength of the transition(s) under
investigation), and the analytical tools that have been used (e.g. model
atmospheres). 

Our sample was assembled from the literature, following some selection
criteria on the quality of the analyses, i.e. high resolution and high S/N.
Since the Li~I line falls in a very clean spectral region, with very few
neighboring absorption lines, the placement of the continuum is usually
quite accurate (on the order of 1-2\%) if the data quality is high. This
uncertainty is usually included in the uncertainty associated to the
equivalent width measurement.

For the latter, because of the generally quite satisfactory agreement
between different literature sources (on a given target, cf Fig.~1) we
decided to use as our final EW the arithmetic mean of all the measurements,
and take the dispersion around the mean as the uncertainty on each
measurement. When only one measurement was available, the associated
uncertainty is the error quoted in the original work. Table\,6 reports both
the error on each EW and the corresponding 1$\sigma$ uncertainty on the Li
abundance. Except for few cases, the latter are well below 0.1~dex. The
uncertainty due to a $\pm$3\%\,error in the log{\it gf} value is
$\pm$0.013~dex.  

Lithium abundances are known to be very sensitive to the chosen effective
temperature, but their dependence on the other parameters, i.e. gravity,
metallicity, and microturbulence is negligible. Common uncertainties on
\logg, [Fe/H] and $\xi$ ($\pm$0.25~dex, $\pm$0.15~dex, and $\pm$0.3~\kms\,
respectively) affect the final Li abundances by at most 0.005~dex, 0.015~dex,
and 0.003~dex. When summed under quadrature, the resulting uncertainty is
around 0.017~dex only.   

On the contrary, the dependence of Li abundances on the effective temperature
is much stronger. An uncertainty of $\pm$ 70\,K in \teff\ (commonly quoted as
a reasonable uncertainty on this parameter) translates into a $\pm$0.05\,dex
on the lithium abundance. For this work, we have considered only the
uncertainties associated to the photometric indices (b-y) (generally quoted
to be around 0.008~mag, cf Nissen et al. 2002) and $\beta$ from which we have
derived our reddening estimates (generally quoted to be around 0.011~mag).
When summed under quadrature, this gives us an average uncertainty on our
effective temperatures of $\pm$75~K, which corresponds to $\pm$0.054~dex in
A(Li). 

Combining all the uncertainties together, we find that depending on the
$\pm1\sigma$(EW) error on A(Li) our best achievable accuracy is 0.06~dex.
In the worst cases it could be as high as 0.15~dex, but one should notice
that for all the stars for which a Li abundance uncertainty larger than
0.1~dex has been derived, the equivalent width of the Li~I line is always
quite small (in the 5-15~m\AA\,range) with a very significant 1$\sigma$ EW
quoted error. Although we do not have the observed spectra available for
further checking, this indicates that S/N ratios on the order of 100 are
probably too low for accurate measurements of weak Li lines. If one were
to exclude those stars with very small equivalent widths (and large quoted
uncertainties) then our final (individual) accuracies range between 0.06
and 0.1~dex.
 
Last but not least, one should not forget that our abundances were
derived based on Kurucz non-overshooting model atmospheres and that most
of the current Li analyses are carried out under LTE assumptions (with NLTE
corrections applied to them) and with one-dimensional model atmospheres.
Choosing a different treatment of the convective motions in the model
atmospheres (i.e. choosing the so-called Kurucz overshooting models) has
the effect of deriving slightly higher Li abundances (by $\simeq$0.08~dex)
but both sets of models carry similar uncertainties which are difficult to
quantitatively assess. NLTE corrections also carry their own uncertainty,
but this is small and well within the average abundance errors, according
to Carlsson et al. (1994). Finally, although Li abundances determined using
3D-hydrodynamical model atmospheres and corrected for NLTE effects differ
from 1D NLTE Li abundances by less than 0.1~dex (0.05~dex for the few stars
that have been investigated so far, cf Asplund et al. 2003 and Barklem et
al. 2003), these same authors warn about possible dependences on temperature
and metallicity, that could clearly affect any discussion on the existence
or lack thereof of a slope in the A(Li)-plateau with \teff\, and/or
metallicity. Hence, no conclusion can be final, until 3D NLTE effects on
Li abundances are mapped on a larger stellar parameters space. 

\subsection{More on the accuracy issue}

Since one of the main focuses of this work is a critical assessment of how
accurately Li abundances in halo stars can be determined and the Li plateau
can be characterised (via its width, spread and slopes with \teff\,and
metallicity), comparisons between our derived Li abundances and previous
analyses are not very significant, especially since our analysis is not
based on newly observed spectra. Our work does not aim at showing that,
with our consistently determined temperature scale, we can now better
describe the properties of the A(Li)-plateau. On the contrary, our analysis
has so far pointed out that although this is clearly a must, even with such
a careful determination of the temperature scale, many uncertainties remain
especially for large samples. 

However, since the opposite findings by some earlier analyses (e.g. R96, R99,
and BM97) have indeed been among the initial triggers of this work, we think
it is useful to further comment on few points. 

First of all, we note that most of the discrepancies originally present among
some stars common to R96, R99, and BM97 (which some of the earlier claims for
dispersion and/or slopes may have originated from) can be fully explained
by differences in the stellar parameters adopted during the various analyses.
Table\,7 summarises these comparisons for some objects common to these works,
with columns 2 to 5 reporting the differences in \teff, metallicity, EW (if
any), and the A(Li) abundance as reported by the original investigators
(always given as ``R96$-$BM97''). When, for a given target, a double entry
is present, this second row of values corresponds to ``R99$-$BM97''. The
last column of the Table reports the remaining difference in the final A(Li)
value, after having taken into account the differences in \teff, [Fe/H], and
EW. 
 
\begin{table}[t]
\label{table6}
\footnotesize{
\caption{Discrepant cases in the original works of R96, R99, and BM97.
Columns identified by a $\Delta$ symbol represent the difference on that
given quantity (\teff, [Fe/H], EW, and lithium abundance) between R96 and
BM97 (when a target has a second entry, this refers to R99 - BM97)} 
\begin{flushleft}
\begin{tabular}{rrcccc}
\hline
\hline
\noalign{\smallskip}
HIP    & $\Delta$\teff & $\Delta$[Fe/H] & $\Delta$EW & $\Delta$A(Li) &
$\Delta$A(Li)$_{end}^{\mathrm{a}}$ \\
       & K             & dex            & m\AA       & dex           &
dex                                \\ 
\noalign{\smallskip}
\hline
3430   & +144 & -0.29 &  0.0 & +0.09 & -0.02 \\
11952  &  -94 & -0.11 &  0.0 & -0.08 & -0.01 \\
12807  & -207 & +0.51 &  0.0 & -0.11 & -0.03 \\
       & -287 & +0.58 & +5.9 & -0.11 & -0.05 \\
14594  & -200 & +0.05 &  0.0 & -0.18 & -0.04 \\
23344  & +230 & -0.90 &  0.0 & +0.15 &  0.01 \\
       & +130 & -0.77 & -1.5 & +0.05 &  0.02 \\
42592  & -164 & +0.20 &  0.0 & -0.12 & -0.01 \\
       & -186 & -0.02 & -2.9 & -0.18 &  0.01 \\
44605  & -240 & -0.34 &  0.0 & -0.23 & -0.05 \\
66673  & -178 & -0.15 &  0.0 & -0.10 &  0.03 \\
       & -248 & +0.18 & -5.8 & -0.25 &  0.06 \\
68592  & -142 & -0.49 &  0.0 & -0.09 &  0.02 \\
       & -192 & -0.72 & +3.2 & -0.01 &  0.07 \\
78640  & -179 & +0.17 &  0.0 & -0.17 & -0.05 \\
87693  & -361 & -0.25 &  0.0 & -0.25 &  0.01 \\
96115  & -160 & +0.18 &  0.0 & -0.11 &  0.00 \\
       & -160 & +0.04 & +1.6 & -0.11 & -0.03 \\
114962 & -142 & +0.23 &  0.0 & -0.15 & -0.06 \\
\noalign{\smallskip}
\hline
\end{tabular}
\end{flushleft}
\begin{list}{}{}  
\item[$^{\mathrm{a}}$]: $\Delta$A(Li)$_{end}$ represents the remaining
discrepancy in the lithium abundance after having taken into account the
differences in \teff (col.~2), [Fe/H] (col.~3), and EW (col.~4). 
\end{list}
}
\end{table}

This is indeed very positive, but in general we find a little worrysome the
revisions made by R99 to some of their measured equivalent widths, compared
to their own previous analysis from 1996. Although both analyses are based
on high quality, high S/N spectra, EW measurements of the \LiI\,line for the
same star differ by as much as 6~m\AA\, which for the \teff\, range covered by
the R99 sample correspond to 0.06~dex in lithium abundance. 

Also, the effective temperatures of the 10
stars common to the analyses of R99, BM97, and this work span respectively
220~K, 358~K, 520~K. In other words, the data-sample that Ryan et al.
carefully chose to span a very narrow range of effective temperatures (and
that indeed did so on their \teff\,scale), it is found to cover a much larger
interval when the temperature is derived following different prescriptions.
And all three analyses used self-consistent methods! 

Another example stressing the weakness of the temperature scale issue
comes from the comparison of our results with the recent work by Mel\'endez \&
Ram\'{\i}rez (2004). The latter have carried out a study similar to ours, in
which 62 halo dwarfs (of which, in the end, 41 were used to discuss the mean
Li abundance of the plateau) were analysed based on a newly derived and
improved IRFM-based \teff\,scale (Ram\'{\i}rez \& Mel\'endez 2005a,b) and
using Li equivalent width measurements available in the literature. 
A quick check between the effective temperatures reported in their Table~1
(Mel\'endez \& Ram\'{\i}rez 2004) and our values, for the 32 stars we have
in common, shows both an offset (their temperature scale is hotter) and a
larger temperature interval spanned (737~K versus 583~K, the latter from
our analysis). The offset implies that their mean Li-plateau value will be
higher than ours. The fact that their temperature scale
is hotter than ours is not in contradiction with what is shown in the bottom
panel of Fig.~11, where a much larger sample of stars is plotted.

\begin{figure}[t]
\rotatebox{0}{\resizebox{9cm}{!}{\includegraphics{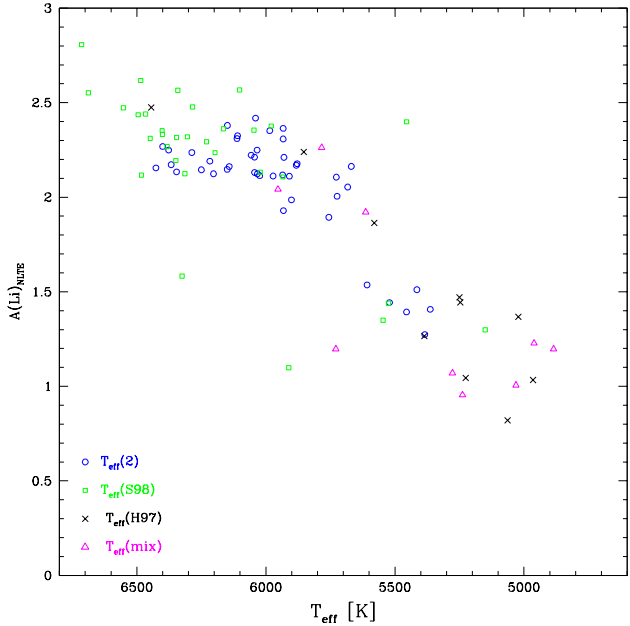}}}
\caption{Lithium abundances as determined by using the sets of temperatures
\teff(2) and \teff(S98) for the {\sl clean} sample (half-half, circle and
square symbols respectively), \teff(H97) and \teff(mix) for the {\sl $\beta$}
and {\sl ubvy} sample (crosses and triangles respectively), in the
attempt of mimicking a common situation in which reddenings are taken from
different sources}
\label{Eby_mix}
\end{figure}

Still related to the determination of an accurate temperature scale, and as
already seen in \S~5, reddening plays an important role. For instance,
R99 noted that their reddening estimates based on two different methods
(Str\"omgren photometry and reddening maps) showed a clear discrepancy of
about 0.02~mag, with the Str\"omgren-based E(b-y) values being higher despite
the expected relationship E(B-V)$\simeq$1.35E(b-y). The solution chosen by
these authors was to give higher weight to the maps-based reddening values
based on the consideration that their targets were bright, hence a low
intrinsic reddening might be expected. Therefore, they systematically lowered
all the Str\"omgren-based E(b-y) values by 0.02~mag before averaging the two
methods. As we do not know the final answer either, we cannot say if this is
a good solution or not. At least for those objects we have in common these stars 
fall well outside the inner 50-70pc of the solar neighborhood 
(they span distances up to 1kpc), for which a low intrinsic reddening could be questionable. 

Even more instructive is to evaluate the realistic situation in which one
may refold on selecting reddening values from different sources, such as
InfraRed dust maps and/or from the literature, because one method alone
cannot be applied homogeneously to the entire sample under investigation.
Keeping in mind that the exact details of such situation are very difficult
to foresee, hence to reproduce, we tried to mimick such case by plotting a
mix of lithium abundances: 
Fig.~\ref{Eby_mix} shows A(Li)$_{NLTE}$ versus \teff, where
the lithium abundances were derived by selecting \teff(2) and \teff(S98) for
the {\sl clean} sample (half-half), and \teff(H97) and \teff(mix) for the
$\beta$ and {\sl ubvy} samples respectively. This likely represents an
extreme case, but it certainly gives an idea of what effect could be expected.
Also, please note that the A(Li) values corresponding to \teff(mix) were
computed only for those stars for which this solution was applied (cf \S~5.1
and \S~5.2).

In summary, despite the seeming convergence at least on the
absence of dispersion, the finding of discordant results is not surprising 
if some (or all) of the above-mentioned points are kept
in mind. At the moment the only claim for a tilted A(Li)$_{NLTE}$-plateau
is with metallicity, but most ``metallicities''
quoted in the literature are still derived from neutral iron lines, the
formation of which is subject to NLTE conditions. Furthermore, R99 (as well
as our work) have used metallicity values extracted from a careful inspection
of the literature, which does not guarantee the homogeneity required for
discussing the A(Li) {\sl vs} [Fe/H] trend. Since the whole discussion of
a possible slope of the lithium plateau with metallicity is centred on a
small dependence, we encourage future analyses of Li abundances to determine
the metallicities spectroscopically, possibly from Fe~II lines (insensitive
to NLTE effects).

\begin{figure}[t]
\rotatebox{0}{\resizebox{9cm}{!}{\includegraphics{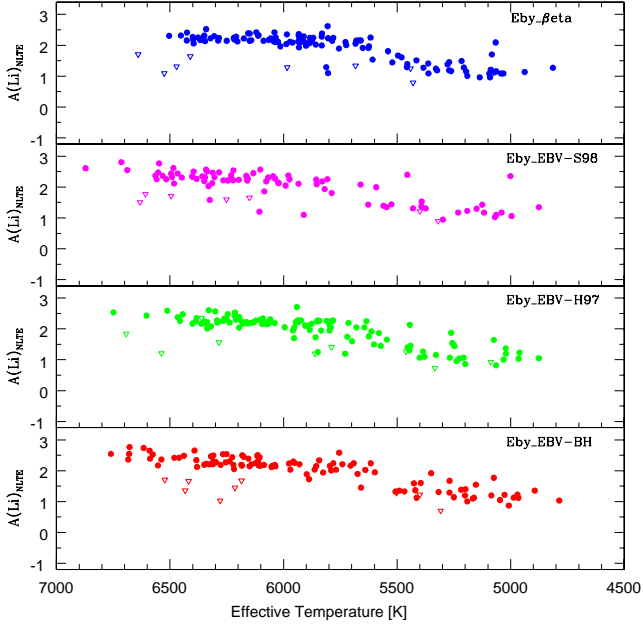}}}
\caption{Lithium abundances computed with four different sets of temperatures
as derived by using different reddening estimates (as reported in the upper
right corner of each pan). Upside-down open triangles represent abundance
upper limits.}
\label{Li4Teff}
\end{figure}

\section{Preliminary results on the mean lithium abundance and the dispersion}

\subsection{Last words of caution before further analysis}

A first visual comparison of the different A(Li)$_{NLTE}$ abundances we
have derived for each star is presented in Fig.~\ref{Li4Teff}, where the first 
three (from the top) panels show how differently the 
plateau may appear when the lithium abundance has been derived from a
different set of temperatures. The bottom pan is shown only for comparison
purposes, and represents the A(Li)$_{NLTE}$ abundances as derived from
photometry that has been dereddened using the BH {\ion{H}{i}} maps. 
Figure~\ref{Li4Teff} shows how claims of dispersion or lack thereof 
from the same data-sample are perfectly plausible 
(see also our discussion in \S~7).

\begin{figure}[t]
\centerline{
\psfig{figure=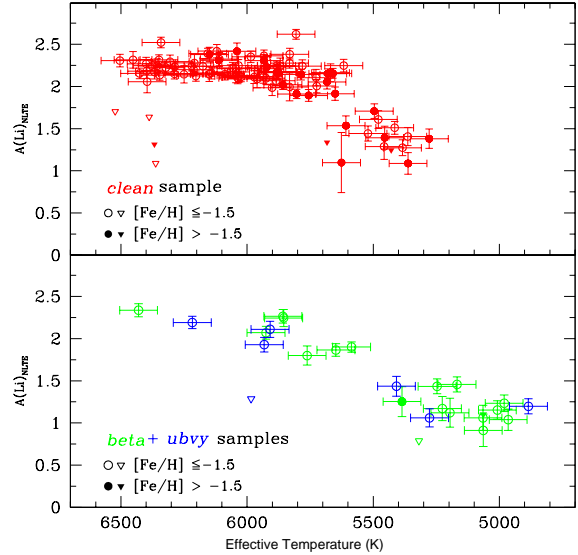,height=8cm}
}
\caption{Final lithium abundances with associated 1$\sigma$-errorbars as a
function of effective temperature (using \teff(2)) are shown for our {\sl
clean} sample (top) and {\sl beta $+$ ubvy} samples (bottom)
\label{ALinlte_vsTeff}
}
\end{figure}

\begin{figure}[t]
\centerline{
\psfig{figure=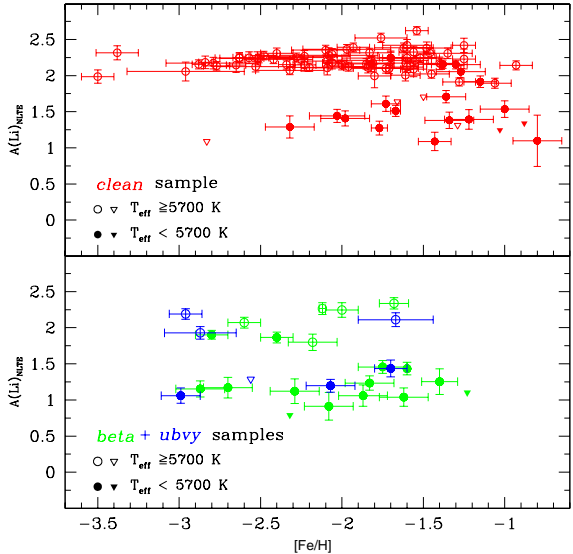,height=8cm}
}
\caption{Final lithium abundances with associated 1$\sigma$-errorbars as a
function of metallicity (using \teff(2)) are shown for our {\sl clean} sample
(top) and for our {\sl beta $+$ ubvy} samples (bottom)
\label{ALinlte_vsFesurH}
}
\end{figure}

In addition, since our {\sl complete} sample is a non-homogeneous sample
(because of the compromises made on the derivation of the effective temperature
for some of the stars), any discussion of the width and slope of the
A(Li)-plateau requires the separation of the {\sl complete} sample into
the {\sl clean}, {\sl $\beta$}, and {\sl ubvy} sub-samples. In
Figs.~\ref{ALinlte_vsTeff} and \ref{ALinlte_vsFesurH}, 
we plot the A(Li)$_{NLTE}$ values {\it vs} our 
\teff\,scale (i.e. \teff(2)) and [Fe/H] respectively, with
the {\sl clean} sample always plotted in the top panel and the {\sl
$\beta$} and {\sl ubvy} samples in the lower panel. Moreover, open
symbols refer to objects with [Fe/H]$\leq-$1.5 in 
Fig.~\ref{ALinlte_vsTeff} and \teff$\geq$5700\,K in 
Fig.\ref{ALinlte_vsFesurH}, and filled symbols
represent respectively stars with [Fe/H]$>-$1.5 and \teff$<$5700\,K.
Upside-down triangles always identify abundance upper limits. 

In the rest of the paper, we give our results
regarding the characteristics of the plateau for the {\sl clean} 
sample on one hand, and for the {\sl complete} (i.e., {\sl clean + 
$\beta$ + ubvy}) sample on the other hand. 
Additionally, we will consider as ``plateau stars" those with 
\teff $\geq 5700$~K and [Fe/H]$\leq -1.5$. 
The metallicity limit is taken in order to avoid any contamination
by lithium production from the various possible stellar sources
(e.g., Travaglio et al. 2001).
The cutoff in \teff is chosen for comparison reasons with 
previous analyses in the literature. However for the purposes of 
constraining stellar evolution models we will also discuss the cases  
where the lower limit in \teff is increased to 6000~K 
in order to avoid proto-stellar lithium destruction (in the case 
of the dwarfs, see \S~10) or dilution at the very beginning of 
the first dredge-up phase (in the case of the post-turnoff stars, 
see \S~11). 
Our results will be given considering the stars 
with lithium abundance determinations only, the case of stars 
with upper limits being considered separately in 
\S~13\footnote{For the metallicity and effective temperature range of the 
plateau, only three stars have Li upper limits, namely 
HIP 72561, 81276 and 100682}.

\subsection{The mean lithium plateau abundance}

Considering the stars with [Fe/H]$\leq -1.5$ 
and in the case of the most relaxed \teff\, lower limit (5700~K, cf 
Figs.~\ref{ALinlte_vsTeff} and \ref{ALinlte_vsFesurH}) we obtain 
$$A(Li)_{NLTE} =2.2138 \pm 0.0929$$
and 
$$A(Li)_{NLTE} =2.2080 \pm 0.0947$$
for the {\sl clean} and {\sl complete} samples respectively, with rms values of
0.0587 and 0.0530. 
This is compatible with a normal distribution (i.e., as would be expected 
from the observational errors). 

For the stars with \teff$\geq$6000\,K, we
find a mean lithium abundance of 
$$A(Li)_{NLTE} =2.2243 \pm 0.0748$$ 
for the {\sl clean} sample, and of 
$$A(Li)_{NLTE}=2.2368 \pm 0.0840 $$
for the {\sl complete} sample. 
In both cases the dispersion values are slightly higher than the rms
of the estimated observational error (0.0587), but compatible with no
dispersion on the plateau. 

Similar conclusions on dispersion can be drawn if no
lower limit on [Fe/H] (in our specific case this was set to $-$2.1) is
assumed in the \teff-(b-y)$_o$ calibration (see \S~5 for more details).
The higher effective temperatures thus derived and shown in Fig.~3 would have
the only effect of increasing our mean A(Li) plateau values by
0.025-0.035~dex. 

We would like to note that the 6 stars that could be
moved to the {\sl ubvy} sample (if we were to consider a 3 digit precision
on $\beta$, cf end of \S~5.2) are not relevant in the final
discussion of the Li plateau spread as they are cool stars that have
their lithium already depleted (5 of them) or a normal plateau lithium
abundance (1 star with A(Li)$_{NLTE}$=2.22).

The absence of intrinsic dispersion that we get is in agreement with the
results of Molaro et al. (1995), Spite et al. (1996), Bonifacio \& Molaro
(1997), Ryan et al. (1999) and Mel\'endez \& Ram\'{\i}rez (2004). 

Furthermore, because all these works (ours included) find quite consistent
A(Li) plateau values (note that the apparently higher plateau value found
by Mel\'endez \& Ram\'{\i}rez (2004, A(Li)$_{NLTE}$=2.37) should be corrected
downwards by 0.08~dex because of the different Kurucz model atmospheres
employed in their and our study, making it then in closer agreement with our
findings), it seems that in general the relatively low lithium abundance
(when compared to the CMB+SBBN result) seen in metal poor halo stars is a
very robust result. Assuming the correctness of 
the CMB constraint on the value of the baryon-to-photon ratio we are
then left with the conclusion that the Li abundance seen at the surface of
halo stars is not the pristine one, but that these stars have undergone
surface lithium depletion at some point during their evolution. 

Let us try now to look for some constraints on the depletion mechanism(s). 

\section{Evolutionary status of the stars}
Using the data we have gathered and homogenized in the first part of this
paper, we will now look at the Li plateau by adding one extra dimension
to the problem, namely by considering the evolutionary status of each star
of our {\sl complete} sample. 
Indeed not all our objects are dwarf stars. 
The contamination from evolved stars has thus to be evaluated 
in order to precisely determine the lithium abundance along the 
plateau and to look for the trends and for the depletion factor. 

\subsection{Input quantities}
We use the HIPPARCOS (ESA 1997) trigonometric parallax measurements to
locate precisely our objects in the HR diagram. Among the 
118 stars of our
{\sl complete} sample, 3 objects (HIP 484, 48444, and 55852) have spurious
Hipparcos parallaxes, and are thus rejected from further analysis. 

Intrinsic absolute magnitudes M$_V$ are derived from the m$_V$ and the
parallaxes given in the Hipparcos catalogue. 
We determine the bolometric corrections BC by using the relations between
BC and V-I (these quantities being also taken from the Hipparcos catalogue)
given by Lejeune et al. (1998) and which
are metallicity-dependent. We use the values of [Fe/H] derived in our final
analysis (Table\,6, column 2). We first iterate using the \logg\, values
available in the literature, and finally attribute to our stars the \logg\,
value derived from their position in the HR diagram 
(Table\,9, column 10).
Finally, we compute the stellar luminosity and the associated error from
the one sigma error on the parallax. All the relevant quantities are given
in Tables\,8 and 9\footnote{available in their entirety on-line}.  

\begin{table}
\label{tableevolutionarystatus}
\caption{Quantities extracted from the HIPPARCOS catalogue for our sample stars
((B-V) and E(B-V) are given in Table~4). The whole table is available on-line}
\begin{tabular}{cccccc}
\hline
\hline \\
\multicolumn{1}{c}{HIP}&
\multicolumn{1}{c}{V}&
\multicolumn{1}{c}{Plx}&
\multicolumn{1}{c}{e$\_$Plx}&
\multicolumn{1}{c}{d}&
\multicolumn{1}{c}{VI} \\
\multicolumn{1}{c}{}&
\multicolumn{1}{c}{mag}&
\multicolumn{1}{c}{(mas)}&
\multicolumn{1}{c}{(mas)}&
\multicolumn{1}{c}{(pc)}&
\multicolumn{1}{c}{} \\[1mm]
\hline \\
   911  &  11.80  &  6.13   & 5.67  &  163  &  0.64 \\
  3026  &   9.25  &  9.57   & 1.38  &  104  &  0.54 \\
  3446  &  12.10  & 15.15   & 3.24  &   66  &  0.58 \\
  3564  &  10.60  &  2.07   & 2.16  &  483  &  0.67 \\
 8572  &  10.34  &  3.22   & 1.75  &  310  &  0.50 \\
   \ld  & \ld     & \ld     & \ld   & \ld      &  \ld  \\
[1mm]
\hline \\
\end{tabular}\\
\end{table}

\subsection{Determination of the stellar evolutionary status}

\begin{figure}
\centerline{
\psfig{figure=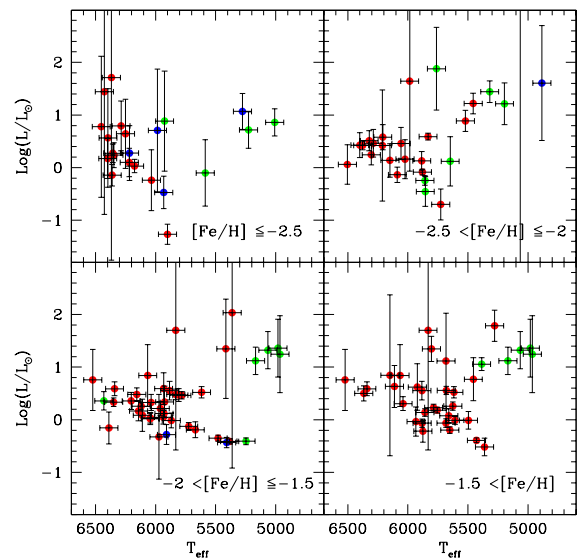,height=8cm}
}
\caption{HR diagram for our {\sl complete} sample stars for separate metallicity bins 
(using \teff(2))
\label{dhr_entiresample}
}
\end{figure}

The resulting HR diagrams are shown in Fig.~\ref{dhr_entiresample} 
for our {\sl complete} sample split in four metallicity bins. 
The evolutionary status of each star has been determined on the basis 
of these HR diagrams, and is given in Table\,9.
Each star has been assigned to one of the following sub-classes: 
5 identifies the dwarfs (i.e., main-sequence stars), 4.5 stands for the
turnoff stars, 4 for the subgiants (i.e., stars crossing the Hertzsprung gap), 
3.5 for the stars at the base of the RGB, and 3 for the stars on the RGB. 
``Post-main sequence stars" can be easily located on the HRD (see
Fig.~\ref{dhr_entiresample}) as those stars with Log(L/L$_{\odot}$) higher
than $\sim$ 0.4. In this luminosity range, we identify as turnoff and subgiant
 stars those with \teff $\geq$ 5600~K, whereas cooler stars are
classified as ``base RGB" or RGB stars depending on their luminosity. This
limit in effective temperature is chosen because it corresponds to the
approximate value where lithium dilution is expected to occur at the very
beginning of the first dredge-up (i.e., Deliyannis et al. 1990, Charbonnel
1995). The distribution in these sub-classes is given in Table~10. For most
of our objects, the classification was performed unambiguously. Some stars
(21 in total) were first classified as ``unsure", the uncertainty being due
to the large errorbar on the derived luminosity inherited from the error on
the Hipparcos parallax (see Fig.~\ref{dhr_entiresample}). However, by
cross-checking the luminosity and the corresponding gravity obtained from
Hipparcos data with the M$_V$ and gravity values quoted in the
literature, we were able to attribute a relatively certain evolutionary status
(given in the last row of 
Table~10) to all these objects, thus keeping them
in our statistical analysis. 
As a result, the {\sl clean} sample (respectively the {\sl complete} sample)
contains 
49 (59) dwarfs, 5 (5) turnoff stars, 29 (31) subgiants, 
5 (18) base RGB stars and 2 (5) RGB stars. 

\begin{table*}
\label{tableevolutionarystatus}
\footnotesize{
\caption{Characteristics and evolutionary status of the sample stars.
The whole table is available on-line}
\begin{tabular}{cccccccccc}
\hline
\hline \\[0.4mm]
\multicolumn{1}{c}{HIP}&
\multicolumn{1}{c}{M$_V$}&
\multicolumn{1}{c}{M$_V$}&
\multicolumn{1}{c}{BC}&
\multicolumn{1}{c}{Mbol}&
\multicolumn{1}{c}{Log(${{L}\over{L_{\odot}}})$}&
\multicolumn{1}{c}{Log(${{L}\over{L_{\odot}}})$}&
\multicolumn{1}{c}{e$\_$Log(${{L}\over{L_{\odot}}})$}&
\multicolumn{1}{c}{log~g}&
\multicolumn{1}{c}{status$^s$} \\
\multicolumn{1}{c}{}&
\multicolumn{1}{c}{}&
\multicolumn{1}{c}{dered}&
\multicolumn{1}{c}{}&
\multicolumn{1}{c}{}&
\multicolumn{1}{c}{}&
\multicolumn{1}{c}{dered}&
\multicolumn{1}{c}{}&
\multicolumn{1}{c}{}&
\multicolumn{1}{c}{} \\[1mm]
\hline \\
   911  &   5.74  &  5.74  &  -0.176  &  5.56 &  -0.32 &  -0.32  &  0.80 & 4.50 & 5  \\
  3026  &   4.15  &  4.15  &  -0.169  &  3.99 &   0.31 &   0.31  &  0.13 & 3.85 & 5  \\
  3446  &   8.00  &  8.00  &  -0.085  &  7.92 &  -1.27 &  -1.27  &  0.19 & 4.50 & 5 \\
  3564  &   2.18  &  2.06  &  -0.221  &  1.96 &   1.12 &   1.17  &  0.91 & 3.50 & 4 \\
  8572  &   2.88  &  2.79  &  -0.112  &  2.77 &   0.79 &   0.83  &  0.47 & 3.85 & 4 \\
   \ld  & \ld     & \ld    & \ld      & \ld   & \ld      & \ld       & \ld     & \ld  & \ld \\
[1mm]
\hline \\
\end{tabular}\\
\begin{tabular}{llll}
(s) Status : 5:dwarf - 4.5:turnoff - 4:subgiant - 3.5:base RGB - 3:RGB \\
(see the text for more details on the adopted definitions of these statuses) \\
\end{tabular}
}
\end{table*}

\begin{figure}
\centerline{
\psfig{figure=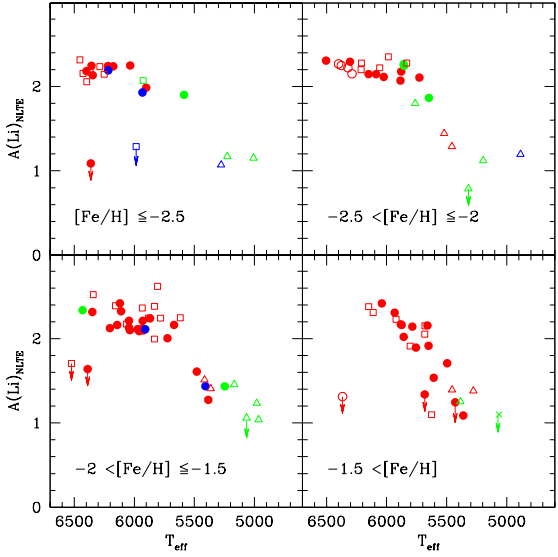,height=8cm}
}
\caption{A(Li)$_{NLTE}$ versus \teff(2) for our {\sl complete} sample
stars, for separate metallicity bins. 
The different symbols indicate the evolutionary status of the stars. 
Filled circles : dwarfs. Open circles : turnoff stars. Open squares :
subgiants. Open triangles : stars at the base of the RGB. Crosses : RGB stars. 
The arrows indicate the lithium upper limits
\label{ALi_vsTeff_vsFesurH_vsstatus}
}
\end{figure}

In Fig.~\ref{ALi_vsTeff_vsFesurH_vsstatus} 
we plot A(Li)$_{NLTE}$ vs \teff for four metallicity bins 
and indicate the evolutionary status of the stars by different symbols.
It is interesting to note that for [Fe/H]$\leq-$2, all the coolest stars 
(i.e., with \teff $\leq$ 5500~K) are actually evolved stars
(see also Fig.~\ref{ALi_vsTeff_vsFesurH_naines}). 
This fact must be taken into account when comparing stellar evolution 
models and observations.
In the [Fe/H] range between $-$2 and $-$1.5, the decrease of lithium 
relative to the plateau dwarf stars appears at $\sim$ 5600~-~5700\,K.

\subsection{Comparison with previous work}

In their careful analysis of the classical mechanisms\footnote{Nuclear burning 
at the basis of the convective envelope on the pre- and early-main sequence, 
diffusion by gravitational settling on the main sequence and at the
turnoff, convective dredge-up on the early post-main sequence, and dilution
on the post-main sequence} that could alter the
surface Li abundance of halo stars at different phases of their evolution, 
Deliyannis et al. (1990, hereafter D90) focused on the
separation of the halo population into pre- and post-turnoff groups as
an essential prerequisite to understand the Li observations. 
At that time they considered a limited sample of halo stars. They used the 
trigonometric parallaxes and V-colors from the Yale Parallax Catalogue
(van Altena et al. 1989) when available to determine M$_V$; otherwise M$_V$ 
was taken from the original observational papers. 
We have 36 objects in common with D90's sample.
For 19 of them the same evolutionary status has been attributed in both D90
and our study. However 13 of the stars which were identified as ``pre-turnoff"
objects by D90 appear to be more evolved stars on the basis of their more
precise and reliable Hipparcos parallaxes. 
Conversely 4 out of the 7 stars which were claimed to be ``post-turnoff"
objects in D90's study are dwarf stars. 

Ryan \& Deliyannis (1998) took advantage of the Hipparcos parallaxes (when
available) in order to separate their sample of halo stars cooler than the
plateau into dwarf and subgiant classes. The 14 stars that we have in common
with their sample have been attributed the same evolutionary status in both
analyses (theirs and ours). 

Note that our definition of a subgiant is different and more strict, as far
as the effective temperature range is concerned, than the one adopted by
PSB93, 
who studied the lithium abundances for 79 so-called halo subgiants.
PSB93 constructed their sample on the basis of {\it ubvy} photometry (more
precisely from their location in the {\it c$_1$} vs {\it (b-y)} plane) from
several catalogs of metal-poor stars. Hipparcos parallaxes are now available
for almost all their objects, allowing a more precise determination of their
location in the HRD. In the present study we did not consider the most evolved
stars of PSB93's sample, i.e., those with \teff below 4800~K. These
objects indeed experience some extra-mixing beyond the first dredge-up and
lose then all the information about their initial lithium content (see
Charbonnel 1995; Weiss \& Charbonnel 2004). They are thus of no help in
discussing the initial lithium content.

\begin{table}
\caption{Distribution of the evolutionary status in our different sub-classes. 
The last line indicates the status that we could attribute to the stars which 
were first labelled as ``unsure" (see the text for more details)}
\begin{center}
\begin{tabular}{ccc}
\hline
\hline
\multicolumn{1}{c}{}
& \multicolumn{2}{c}{Sample} \\
\hline
\multicolumn{1}{c}{}
& \multicolumn{1}{c}{{\sl Clean}}
& \multicolumn{1}{c}{Polluted ({\sl beta}+{\sl ubvy})} \\
\hline
total    & 90 & 27(20+7) \\ 
dwarfs   & 44 & 10(6+4) \\
turnoff  &  3 & 0 \\  
subgiant & 18 & 2(1+1) \\
base RGB &  5 & 13(11+2) \\
RGB      &  1 &  0 \\
unsure   & 19 & 2 (2+0) \\
         & [5dw, 2to, 11sg, 1rgb] & [1dw, 1rgb] \\
\hline
\end{tabular}
\end{center}
\end{table}

\section{The lithium abundance along the plateau for the dwarf stars}

We now concentrate our analysis on the dwarf stars. 
Their lithium abundance is plotted in Fig.~\ref{ALi_vsTeff_vsFesurH_naines}. 

\subsection{The analytical method}

\begin{figure}
\centerline{
\psfig{figure=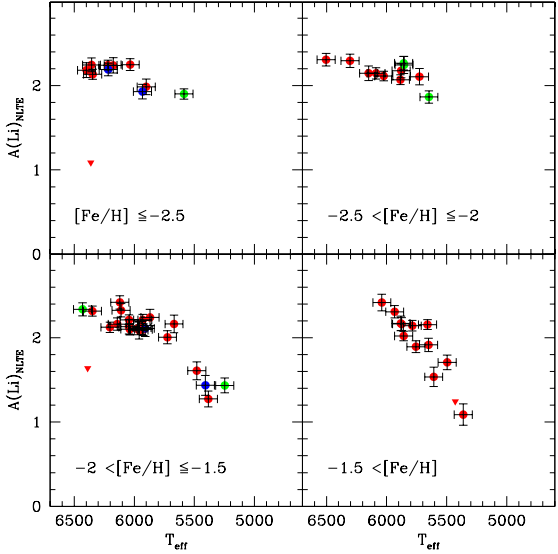,height=8cm}
}
\caption{A(Li)$_{NLTE}$ versus \teff(2) for the dwarfs of our {\sl complete} sample. 
Upside-down triangles represent lithium upper limits
\label{ALi_vsTeff_vsFesurH_naines}
}
\end{figure}

For the reasons given in \S~8.1, we identify as ``plateau stars" those with
[Fe/H]$\leq -1.5$ and with \teff higher than 5700 or 6000~K. The
corresponding numbers of stars in each bin is given in
Table~11. 
For the time being, we keep in our analysis
the possible binary stars (but see \S~12), but we eliminate the stars with
lithium upper limits, that will be discussed separetely in \S~13. We note
that in the metallicity and effective temperature ranges we have chosen to
define the plateau only two dwarf stars of the {\sl clean} sample have Li
upper limits, namely HIP 72561 and 100682. With A(Li)$_{NLTE}$ $\leq$ 1.639
and $\leq$ 1.088 respectively (with [Fe/H] = $-$1.66 and $-$2.83, \teff 
= 6388 and 6362~K), both stars fall well below the plateau (see
Fig.~\ref{ALi_vsTeff_vsFesurH_naines}). 

\begin{table}
\caption{Number of dwarfs in each subsample}
\begin{center}
\begin{tabular}{cccccc}
\hline
\hline
\multicolumn{1}{c}{[Fe/H]}
& \multicolumn{1}{c}{{\sl Clean}}
& \multicolumn{1}{c}{Polluted}
& \multicolumn{1}{c}{{\sl Clean}}
& \multicolumn{1}{c}{Polluted} \\
\multicolumn{1}{c}{}
& \multicolumn{1}{c}{}
& \multicolumn{1}{c}{({\sl beta}+{\sl ubvy})} 
& \multicolumn{1}{c}{} 
& \multicolumn{1}{c}{({\sl beta}+{\sl ubvy})} \\
\multicolumn{1}{c}{}
& \multicolumn{1}{c}{$>$5700}
& \multicolumn{1}{c}{$>$5700}
& \multicolumn{1}{c}{$>$6000}
& \multicolumn{1}{c}{$>$6000} \\
\hline
$\leq$-2.5   &  8 & 0+2  & 7 & 0+1  \\
-2.5 to -2.0 &  9 & 2+0  & 6 & 0+0  \\
-2.0 to -1.5 & 16 & 1+1  & 9 & 1+0  \\
\hline
\end{tabular}
\end{center}
\end{table}

In order to investigate the existence of trends in the
[A(Li)$_{NLTE}$, \teff] and [A(Li)$_{NLTE}$,[Fe/H]] planes  
we performed univariate fits by means of four estimators : 
(1) The least-squares fit with errors in the independant variable only 
(routine FIT of Press et al. 1982);
(2) The least-squares fit with errors in both variables 
(routine FITEXY of Press et al. 1982); 
(3) The BCES (bivariate correlated errors and intrinsic scatter) of 
Akritas \& Bershady (1996);
(4) BCES simulations bootstrap based on 10000 samples. 
All univariate fits in the A(Li)$_{NLTE}$ versus \teff plane
were computed considering an error on \teff of both 75 and 100~K.

\subsection{Mean value and dispersion}

\begin{table}[ht]
\label{tabledispersion}
\caption{Average value of A(Li)$_{NLTE}$, average and standard deviation, and rms for the dwarf stars :
mean$^{\pm {\rm aver}}_{\pm {\rm stand}}$ (rms). We separate the {\sl clean} and the {\sl complete} sample 
and give the results for 2 limits in effective temperature}
\begin{center}
\begin{tabular}{cll}
\hline
\hline
\multicolumn{1}{c}{[Fe/H]}
& \multicolumn{2}{c}{\teff $\geq$5700~K} \\
\multicolumn{1}{c}{}
& \multicolumn{1}{c}{{\sl clean}}
& \multicolumn{1}{c}{{\sl complete}} \\
\hline
$\leq$-2.5 & 2.183$^{\pm 0.070}_{\pm 0.097}$ (0.042) & 2.156$^{\pm 0.093}_{\pm 0.119}$ (0.052) \\
-2.5 to -2.0 & 2.170$^{\pm 0.067}_{\pm 0.087}$ (0.047) & 2.187$^{\pm 0.072}_{\pm 0.085}$ (0.057) \\
-2.0 to -1.5 & 2.178$^{\pm 0.088}_{\pm 0.109}$ (0.042) & 2.183$^{\pm 0.092}_{\pm 0.111}$ (0.067) \\
$\leq$-1.5 & 2.177$^{\pm 0.071}_{\pm 0.098}$ (0.047) & 2.177$^{\pm 0.084}_{\pm 0.104}$ (0.052) \\
\hline
\hline
\multicolumn{1}{c}{}
& \multicolumn{2}{c}{\teff $\geq$6000~K} \\
\multicolumn{1}{c}{}
& \multicolumn{1}{c}{{\sl clean}}
& \multicolumn{1}{c}{{\sl complete}} \\
\hline
$\leq$-2.5 & 2.216$^{\pm 0.038}_{\pm 0.047}$ (0.042) & 2.212$^{\pm 0.037}_{\pm 0.044}$ (0.052)\\
-2.5 to -2.0 & 2.202$^{\pm 0.079}_{\pm 0.091}$ (0.047) & 2.202$^{\pm 0.079}_{\pm 0.091}$ (0.047) \\
-2.0 to -1.5 & 2.225$^{\pm 0.088}_{\pm 0.108}$ (0.042) & 2.236 $^{\pm 0.100}_{\pm 0.115}$(0.067) \\
$\leq$-1.5 & 2.215$^{\pm 0.074}_{\pm 0.088}$ (0.047) & 2.200$^{\pm 0.074}_{\pm 0.088}$ (0.052)\\
\hline
\end{tabular}
\end{center}
\end{table}

The average value of the lithium abundance is given in
Table~12, together with the average, standard deviation
and root mean square of the estimated observational errors for several
subsets of data (in terms of metallicity and effective temperature intervals, 
as well as in terms of {\sl clean} vs {\sl complete} samples) 
of the dwarf sample. 

If we consider the entire metallicity range [Fe/H]$\leq-$1.5, 
the straight average value of the lithium abundance for the plateau dwarf stars 
with \teff $\geq 5700$\,K is 
$$A(Li)_{NLTE} = 2.1768 \pm 0.0711$$
and 
$$A(Li)_{NLTE} = 2.1773 \pm 0.0840$$ 
for the {\sl clean} and {\sl complete} samples respectively. 
The average dispersion is small, but slightly higher than
the respective rms of 0.0474 and 0.0516.

When we restrict our analysis to the dwarf stars with \teff $\geq 6000$~K, the mean value 
increases to 
$$A(Li)_{NLTE} = 2.2154 \pm 0.0737$$ 
and 
$$A(Li)_{NLTE} = 2.2200 \pm 0.0740$$ 
for the {\sl clean} and the {\sl complete} samples respectively. 
Again the average dispersion is small but slightly higher than the rms
(0.0474 and 0.0516 respectively). 

We thus find no evidence of an intrinsic dispersion in the Li abundances along
the plateau. As expected, this result which was already presented in \S~8.2, 
is now reinforced having eliminated the ``pollution" by evolved stars and
focussed on the dwarf stars only. 

As we already mentioned in \S~8.2, the assumption of a lower limit 
on [Fe/H] in our \teff-(b-y)$_o$ calibration does not affect our 
conclusions on dispersion. The only effect when one does not consider this limit 
is to increase the mean A(Li) value of the dwarf sample by 
less than 0.02~dex for the entire metallicity range. If we consider 
only our most metal-poor subsamples the increase in the mean A(Li) value 
is of course slightly higher (from 0.02 up 0.08~dex depending on the 
lowest limit on \teff used for the plateau definition). 

The mean lithium values for the dwarf stars are slightly
lower, although fully compatible within the quoted errors, than the mean
values given in \S~8.2 for the entire sample of stars (i.e., that in which
we did not discriminate the stars according to their evolutionary status). 
This point will be discussed at length in \S~11.

\subsection{The A(Li)$_{NLTE}$ versus \teff correlation}

When we consider the dwarfs with \teff $\geq$ 5700~K and
[Fe/H]$\leq-$1.5,  we find a small slope in the A(Li)$_{NLTE}$-\teff 
plane of $$ 0.019/100~K ~{\rm and}~ 0.028/100~K $$
for the {\sl clean} and {\sl complete} samples respectively. 
These numbers are obtained when we use our standard error on \teff of
75~K. In the more conservative case of an error on \teff of 100~K, we
get respectively 0.026/100~K and 0.033/100~K. 

This is very similar to the slope found by BM97 of
0.02/100~K, and slightly lower than those found by Thorburn (1994) of
0.034/100~K and by R96 of 0.0408/100~K. On the contrary,
Mel\'endez \& Ram\'{\i}rez (2004) do not find any dependence of A(Li) on either
\teff\,or [Fe/H].

In order to focus on the physical processes that may affect the surface
lithium abundance only when the stars are on the main sequence (in other
words, to avoid any pre-main sequence depletion; see e.g. D90, Richard et al.
2004) we made the same computations for the stars with
\teff $\geq$~6000~K. In this case and for the whole [Fe/H]$\leq-$1.5
range, we get a slightly smaller slope, i.e. $$ 0.015/100~K $$ for the {\sl
clean} sample and a very similar slope, i.e. $$ 0.029/100~K $$ for the {\sl
complete} sample, assuming an error on \teff of 75~K. When we consider
an error on \teff of 100~K, we get respectively 0.027/100~K and
0.041/100~K. 

We note that if we split the [Fe/H] interval as done for example in
Fig.~\ref{ALi_vsTeff_vsFesurH_naines}, we find a negative slope of $\sim-$
0.025 for the stars with [Fe/H]$\leq-$2.5, both for the {\sl clean} and {\sl
complete} samples for the stars with \teff $\geq$~6000~K. Small but
positive slopes are detected when we look at the $-$2.5 $<$ [Fe/H]$\leq-$2.0
and $-$2.0 $<$ [Fe/H]$\leq-$1.5 intervals. 

\subsection{The A(Li)$_{NLTE}$ versus [Fe/H] correlation}

\begin{figure}
\centerline{
\psfig{figure=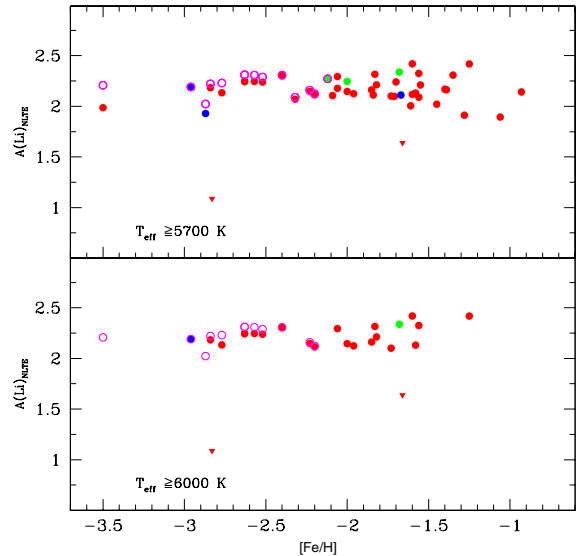,height=8cm}
}
\caption{A(Li)$_{NLTE}$ versus [Fe/H] only for the dwarfs with \teff 
$\geq 5700$~K (upper figure) and with \teff $\geq 6000$~K (lower figure) 
of our entire sample. Upside-down triangles represent abundance upper limits.
The open circles correspond to the case where no lower limit 
on [Fe/H] is assumed in the colour-\teff calibration, while the others 
correspond to the case where this limit is set to -2.1
\label{ALinlte_vsFesurH_naines}
}
\end{figure}

In Fig.~\ref{ALinlte_vsFesurH_naines} we plot A(Li)$_{NLTE}$ {\it vs} [Fe/H]
for the plateau dwarfs of our entire sample but we look for trends 
up to [Fe/H]~=~-1.5 only.  
The highest slope ($\sim$~0.05 up to [Fe/H]~=~-1.5) 
is obtained for the most extended sample,
i.e., for the dwarf stars of the {\sl complete} sample with \teff $\geq$
5700~K, whereas when we consider only the dwarf stars of the {\sl clean}
sample (still with \teff $\geq$ 5700~K) the correlation becomes 
slightly flatter (the slope is $\sim$~0.02). 
When we focus on the dwarf stars with \teff $\geq$ 6000~K, the
dependence of A(Li) with metallicity 
remains small (with a slope of $\sim$~0.03 and 0.02~dex up to [Fe/H]~=~-1.5
for the {\sl complete} and {\sl clean} samples respectively). 
In Fig.~\ref{ALinlte_vsFesurH_naines} we also show the values when
no lower limit on [Fe/H] is assumed in the \teff-(b-y)$_0$ calibration.
A slightly lower slope would have been derived under this assumption.

Our results are intermediate between those of
BM97 ($-$0.05 to 0.00, i.e., no slope) and those of 
R99 (+0.118) and Thorburn (1994; +0.13). We note that the
preliminary results of Asplund et al. (2005) indicate a dependence of
A(Li) on metallicity (the latter computed both as the abundance of iron
and oxygen), characterised by a slope of 0.10. 

We remind the reader that lithium synthesis by galactic cosmic rays (GCR) leads
indeed to an increase of its abundance with metallicity, but this contribution 
safely can be considered small up to [Fe/H]=$-$1.5 (Molaro et al. 1997)
in agreement with our finding.
This can be estimated using the observed correlation between $^9$Be and metallicity
together with the Li/Be ratio predicted by GCR theory (Walker et al. 1985).  

As discussed already in \S 7.2, the trends with
metallicity here derived must be taken with caution, as we have gathered
our metallicities from the literature. This does not guarantee the homogeneity
required for an accurate study of the metallicity dependence. Additionally
such an analysis would require homogeneous spectroscopic
determinations from Fe II lines insensitive to NLTE effects rather than from
neutral iron lines as it is the case for most of the currently available Li
analyses (hence also for our values).
For these reasons we will not continue the discussion nor derive any 
conclusions on the lithium-metallicity relation.  

\subsection{The dwarf stars cooler than the plateau}

As already mentioned in \S~9.2, the current sample does not contain 
cool (i.e., with \teff $\leq$ 5500~K) dwarf stars
at low-metallicity (i.e., [Fe/H]~$\leq-$~2).
This is the first surprise of our check on the evolutionary status
of our sample stars. It thus appears that for the cooler and more 
metal-deficient dwarfs we have no direct clue to the lithium behavior. 
We expect these objects to exhibit the same pattern as that seen in cool open
cluster and Pop I field stars as well as in our sample stars at higher
metallicity (bottom panels of Fig.~\ref{ALi_vsTeff_vsFesurH_naines}),
although this remains to be confirmed observationally.
In the [Fe/H] range between $-$2 and $-$1.5, substantial lithium depletion 
relative to the plateau value sets in at $\sim$ 5600~--~5700\,K.  
Between 5700 and 5100\,K, the lithium depletion slope is 
$\sim$~0.21~dex per 100\,K. 

\section{The lithium abundance in evolved stars}
\begin{figure}
\centerline{
\psfig{figure=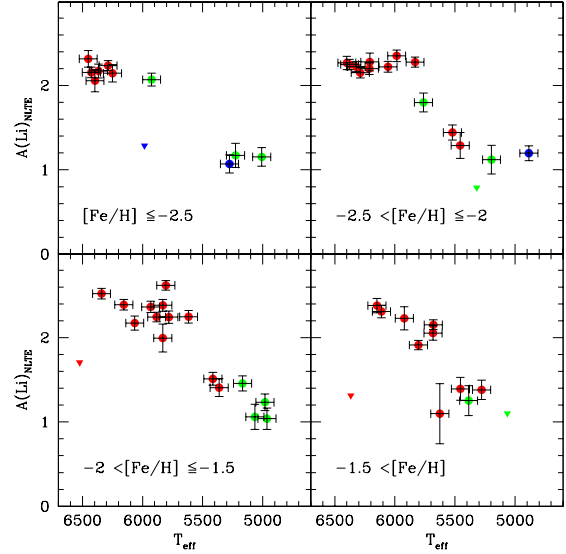,height=8cm}
}
\caption{A(Li)$_{NLTE}$ versus \teff(2) for the turnoff and more evolved
stars of our entire sample for four metallicity bins. 
Triangles are for lithium upper limits. Effective temperature is here
an indicator of the evolutionary status of the stars
\label{ALi_vsTeff_vsFesurH_evolvedstars}
}
\end{figure}

We now address the case of more evolved stars. Their lithium abundance is plotted in
Fig.~\ref{ALi_vsTeff_vsFesurH_evolvedstars} for four metallicity bins as a
function of \teff which is now an indicator of the evolutionary
status. We analyse the lithium behaviour in two separate groups of objects:
turnoff and subgiant stars on one hand, and RGB stars on the other hand. 
We compare our results with the lithium data in globular clusters.

\subsection{The mean lithium value in turnoff and subgiant stars}

\begin{table}
\label{tabledispersion}
\caption{Straight average value of A(Li)$_{NLTE}$, average and standard
deviation for the turnoff and subgiant stars :
mean$^{\pm {\rm aver}}_{\pm {\rm stand}}$ (rms). We separate the {\sl clean} 
and the {\sl complete} samples and present the results for 2
limits in effective temperature}
\begin{center}
\begin{tabular}{cll}
\hline
\hline
\multicolumn{1}{c}{[Fe/H] interval}
& \multicolumn{2}{c}{\teff$\geq$5700~K}\\
\multicolumn{1}{c}{}
& \multicolumn{1}{c}{{\sl clean}}
& \multicolumn{1}{c}{{\sl complete}} \\
\hline
$\leq$-2.5   & 2.180$^{\pm 0.064}_{\pm 0.088}$ (0.093) & 2.164$^{\pm 0.066}_{\pm 0.090}$ (0.053)\\
-2.5 to -2.0 & 2.253$^{\pm 0.045}_{\pm 0.058}$ (0.075) & 2.247$^{\pm 0.042}_{\pm 0.057}$ (0.075) \\
-2.0 to -1.5 & 2.326$^{\pm 0.144}_{\pm 0.187}$ (0.059) & 2.326$^{\pm 0.144}_{\pm 0.187}$ (0.059) \\
$\leq$-1.5 & 2.260$^{\pm 0.098}_{\pm 0.136}$ (0.059) & 2.252$^{\pm 0.099}_{\pm 0.138}$ (0.053) \\
\hline
\hline
\multicolumn{1}{c}{}
& \multicolumn{2}{c}{\teff$\geq$6000~K}\\
\multicolumn{1}{c}{}
& \multicolumn{1}{c}{{\sl clean}}
& \multicolumn{1}{c}{{\sl complete}} \\
\hline
  $\leq$-2.5 & 2.180$^{\pm 0.064}_{\pm 0.088}$ (0.093)  & 2.180$^{\pm 0.064}_{\pm 0.088}$ (0.093)\\
-2.5 to -2.0 & 2.232$^{\pm 0.033}_{\pm 0.046}$ (0.075) & 2.227$^{\pm 0.032}_{\pm 0.043}$ (0.075)\\
-2.0 to -1.5 & 2.362$^{\pm 0.126}_{\pm 0.176}$ (0.059) & 2.362$^{\pm 0.126}_{\pm 0.176}$ (0.059) \\
$\leq$-1.5 & 2.235$^{\pm 0.077}_{\pm 0.109}$ (0.059) & 2.235$^{\pm 0.077}_{\pm 0.109}$ (0.059)\\
\hline
\end{tabular}
\end{center}
\end{table}

As already mentioned in \S~10, we put a lower limit on the effective
temperature (5700 and 6000~K) for our analysis, this time in order to avoid
any lithium depletion due to the first dredge-up. We consider stars both at
the turnoff and on the subgiant branch as defined in \S~9.2.
Two stars are excluded from the analysis, namely HIP 81276 ([Fe/H]~=~$-$1.50,
\teff~=~6523\,K) of the {\sl clean} sample and HIP 83320
([Fe/H]~=~$-$2.56, \teff~=~5984\,K) of the {\sl ubvy} sample, 
because of their A(Li)$_{NLTE}$ upper limits (of 1.705 and 1.287 respectively,
see Fig.~\ref{ALi_vsTeff_vsFesurH_evolvedstars}). 
The average value of A(Li)$_{NLTE}$, the average and standard 
deviation and the root mean square of the observational errors are given
in Table~13 for various metallicity and effective temperature intervals and 
for the {\sl clean} and {\sl complete} samples. 
 
A very surprising result emerges here for the first time: 
Whatever the subsample we consider (i.e., {\sl clean} or {\sl complete}
sample, \teff$\geq$ 5700 or 6000~K), the mean value of A(Li)$_{NLTE}$
always appears to be higher for the subgiant stars than for the dwarfs,
except for the most metal-deficient objects (i.e., with [Fe/H]$\leq-$2.5)
where the mean lithium abundance is very similar in both evolutionary statuses. 

In the case of the evolved stars with \teff $\geq$ 5700~K and
[Fe/H]$\leq-$1.5, 
$$A(Li)_{NLTE} = 2.2599 \pm 0.0997$$
and 
$$2.2524 \pm 0.0990$$ 
for the {\sl clean} and {\sl complete} samples respectively. 
These values have to be compared respectively with 
2.1768$\pm$0.0711 and 2.1773$\pm$0.0840 
for the dwarfs the same range of effective temperature. 

When we restrict our inspection to the stars with \teff $\geq$ 6000~K and
[Fe/H]$\leq-$1.5, the mean value of A(Li)$_{NLTE}$ for the evolved stars
is 
$$2.2349 \pm 0.0769$$ 
for both the {\sl clean} and the {\sl complete} samples\footnote{The {\it ubvy}
 and {\it $\beta$} samples do not contain additional objects}, 
which has to be compared with 
2.2154$\pm$0.0737 and 2.2200$\pm$0.0740 
for the dwarf stars. 

Once again, we note that our conclusions are not affected by having assumed $-$2.1 
as the lower limit on the metallicity
to be used in our colour-\teff\,calibration (see \S~5). 
The only effect of not considering this limit would be a slight increase
of the Li mean value by $\sim$0.04-0.05~dex.

Although the numbers for the dwarf and evolved stars are fully compatible 
within the quoted errors, the difference between the mean lithium values 
of both populations is stricking. 
As already discussed in \S~7.1, the dependence of the lithium abundance on
gravity is weaker than that on effective temperature. At \teff~=~6000~K
 and [Fe/H]~=~$-$1.5, the typical effect is at most +/$-$0.01~dex in A(Li)
for $-$/+1 dex in \logg. We can thus see that even an error of 1~dex on the
attributed gravity (which is very unlikely in view of the good precision of
the Hipparcos parallax for most of the stars) cannot explain the difference
on the mean lithium abundance that we obtain between the dwarf and subgiant
stars. Neither can the dependence of our colour-\teff\,calibration on
\logg. There is indeed a dependence for giant stars because (b-y) measures
the slope of the Paschen continuum, which in turn is affected by a change
downwards of the gravity (cf RM05b). However, what we call ``evolved'' stars
are objects that have just passed the turn-off or that are located on the
subgiant branch: they are not real giants, as defined in RM05b.

\subsection{Dispersion in evolved stars}

Our analysis reaveals another remarkable feature: Post-main sequence Population 
II stars appear to exhibit a non negligible Li dispersion.
As can be seen for example in Fig.\ref{ALi_vsTeff_vsFesurH_evolvedstars},
this is already true at the turnoff and all along the evolution traced by the
effective temperature. 

For the subgiants near 6000~K, PSB93 found a small spread in the 
lithium abundance around a mean value of 2.1. 
We see from the previously quoted numbers that the dispersion is 
actually not negligible in our sample of slightly evolved stars, 
independently of the adopted limit in effective temperature (5700 or 6000~K). 

\begin{figure}
\centerline{
\psfig{figure=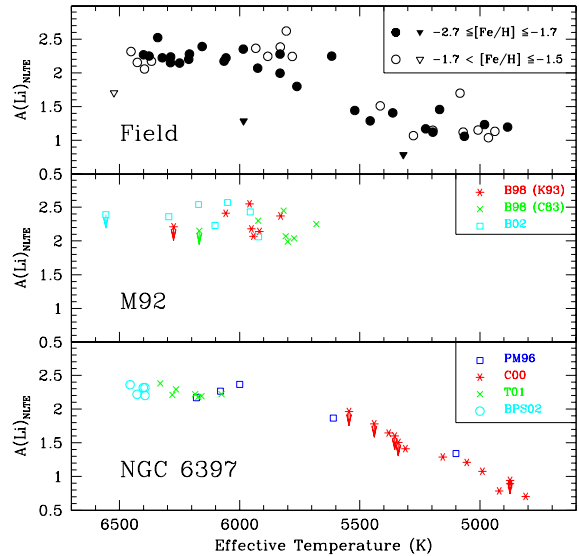,height=8cm}
}
\caption{Lithium abundance versus \teff for the turnoff and more
evolved stars of our sample (upper panel, \teff(2)) and for their counterparts
in the globular clusters M92 and NGC 6397 (lower and middle panel respectively). 
Effective temperature is here an indicator of the evolutionary status of the
stars. For the field stars we focus on our sample stars in a limited 
metallicity range 
around that of the two globular clusters discussed here ([Fe/H]~=~$-$2.52 for
M92, King et al. 1998; [Fe/H]~=~$-$2.02 for NGC 6397, Th\'evenin et al. 2001).
All the Li values are NLTE (see the text). For M92, the original data are from
Boesgaard et al. (1998, B98) and Bonifacio (2002, B02). For B98 study we plot
the data both on the Carney (1983, C83) and King (1993, K93) temperature
scales. The data for NGC 6397 are from Pasquini \& Molaro (1996, PM96), 
Castilho et al. (2000, C00), Th\'evenin et al. (2001, T01) and Bonifacio et
al. (2002, BPS02)
\label{subgiants_field_GC}
}
\end{figure}

For the more evolved (i.e., cooler) stars, 
the lithium abundance is expected to decrease due to the first dredge-up 
at the effective temperature of $\sim$ 5700\,K, this dilution episode being
completed around 5200\,K (i.e., Deliyannis et al. 1990, Charbonnel 1995).
This is what we observe, even though this region is not very well sampled. 
However if all the stars had left the main sequence with the same lithium
content, they should share a common lithium abundance after the first
dredge-up (remember that these stars have approximately the same initial
mass). 
Instead, the evolved stars with \teff lower than
$\sim$ 5500\,K which have already undergone the first dredge-up dilution
exhibit a relatively large lithium dispersion. 

\subsection{Comparison with globular cluster stars}

The question of the lithium dispersion among metal-poor stars has 
already been discussed in the context of globular cluster studies. 
In Fig.~\ref{subgiants_field_GC} we show the data in the only two 
globular clusters for which Li abundances have been reported in stars 
down to the turnoff, namely M92 and NGC 6397 (middle and lower panel
respectively). These are NLTE values, i.e. we corrected the LTE values
reported in the literature for NLTE corrections which were computed
following the same prescriptions we used for our field stars sample.
Unfortunately, the unavailability of Str\"omgren photometry for all
the globular cluster stars analysed in M92 and NGC6397 did not allow
us to re-derive their lithium abundances based on our temperature
scale.
For M92 we rely on two analyses : i) Boesgaard et al. (1998, hereafter B98), 
from which we report the Li values for both temperature scales (Carney 1983,
C83, and King 1993, K93) discussed in their original paper; ii) Bonifacio
(2002, hereafter B02) which is based on B98 equivalent widths and on the 
temperature calibration by Bonifacio et al. (2002, hereafter BSP02). 
For NGC 6397, the Li abundances are taken from Pasquini \& Molaro (1996),
Castilho et al. (2000), Th\'evenin et al. (2001) and Bonifacio et al.
(2002, BSP02).
In Fig.~\ref{subgiants_field_GC} the globular cluster data are compared 
with the NLTE Li abundances of our evolved field stars in the metallicity 
range around that of the clusters ($-$2.7~$\leq$~[Fe/H]~$\leq$~$-$1.7,
black symbols in the upper panel). 

B98 investigated the lithium abundance in seven stars near the turnoff of
the old and metal-poor cluster M92 ([Fe/H]~=~$-$2.52, King et al. 1998).
They reported a dispersion of a factor of $\sim$2.6 for the subgiant stars
in a region around \teff $\sim $~5800~K or 5950~K (depending on the adopted
temperature scale, C83 or K93 respectively) as can be seen in the middle panel
of Fig.~\ref{subgiants_field_GC}. Contrary to B98, B02 concluded that there
is no strong evidence for intrinsic dispersion in Li abundances among the M92
stars, although a dispersion as large as 0.18~dex is possible. B02 actually
warned the reader that no definitive conclusion can be drawn on the intrinsic
dispersion in this stellar cluster on the basis of the currently available
spectra. Better observations with higher {\it S/N} ratios are still needed 
for these very faint M92 stars. 

Although the data points are relatively scarce in this region, one 
sees from Fig.~\ref{subgiants_field_GC} that some dispersion indeed exists
among field stars in the \teff\,range of B98's data. 
The stars HIP 36430, 92775 and 102718 which have respectively 
[Fe/H]~=~$-$2.10, $-$2.18 and $-$1.80,
\teff~=~5985, 
5762 and 5832~K, and very similar
Log(L/L$_{\odot}$)~=~1.69$\pm$1.71, 1.88$\pm$0.79 and 1.71$\pm$2.27,
present significant
differences in their NLTE-Li abundance : 2.352$\pm$0.068, 
1.799$\pm$0.113 and 1.995$\pm$0.165.
This corresponds to the dispersion claimed by B98 in M92. 
In addition at approximatively the same effective temperature the slightly
more metal-poor star HIP 83320 ([Fe/H]~=~$-$2.56, \teff~=~5984~K,
Log(L/L$_{\odot}$)~=~0.77$\pm$1.17) shows only an upper limit for Li
(A(Li)$\leq$1.287)\footnote{We note that the error on the Hipparcos
parallax is relatively high for HIP 36430, 92775 and 83320, which turns into
a significant error bar on the determined luminosities. The status of subgiant
can however be attributed relatively safely to these three objects (see
also Fig.~\ref{dhr_entiresample}).}.  
B98 discussed the case of the halo subgiant BD + 23$^o$3912 (HIP 99423) which 
was found by King et al. (1996) to have a remarkably high lithium abundance 
of 2.56$\pm$0.07. This star is also in our sample, with the following 
parameters : [Fe/H]~=~$-$1.54, \teff~=~5806~K, and
Log(L/L$_{\odot}$)~=~0.48$\pm$0.12. We find a relatively high NLTE Li
abundance of 2.620 for this object. Although its relatively high metallicity
makes this star not very relevant when compared to M92 stars, we think that
it reinforces the case for dispersion. 

The lithium behavior in NGC 6397 ([Fe/H]~=~$-$2.02, Th\'evenin et al. 2001) 
is instead completely different: stars with \teff $\geq 6000$~K 
share the same lithium abundance with essentially no intrinsic
scatter. No lithium abundance has been reported up
to now in this cluster for stars over the $\sim$ 5600 to 6000\,K range. 
Observations in this region would be useful. 

\section{Duplicity and variability}

\begin{figure}
\centerline{
\psfig{figure=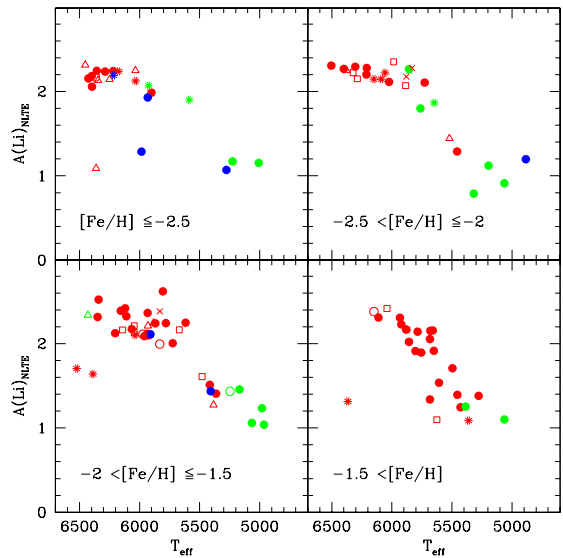,height=8cm}
}
\caption{Duplicity and variability among our {\sl complete} sample. 
Asterisks : confirmed single- or double-lined binaries (Latham et al. 2002, Carney
et al. 2003, 1994). Open squares : stars marked as spectroscopic binaries in
Bonifacio \& Molaro (1997). Open triangles : suspected binaries (Latham et al. 2002). 
Crosses : binaries or stars in double/multiple system as reported in SIMBAD. 
Filled circles : single stars. Open circles : variable stars as reported in
SIMBAD
\label{ALi_vsTeff_vsFesurH_duplicity}
}
\end{figure}

\begin{table}
\caption{Average value of A(Li)$_{NLTE}$ and average deviation 
for [Fe/H]~$\leq$~-1.5 : 
When all the stars are considered regardless of their possible 
duplicity or variability (column a), when all the suspected and confirmed 
binaries are omitted (column b), and when only the confirmed binaries are 
excluded (column c). In cases (b) and (c) the stars reported as variable 
in SIMBAD are not considered.  The first and second lines contain the values 
for the {\it clean} and the {\it complete} samples respectively}
\begin{center}
\begin{tabular}{cccc}
\hline
\hline
\multicolumn{1}{c}{}
& \multicolumn{1}{c}{(a)}
& \multicolumn{1}{c}{(b)}
& \multicolumn{1}{c}{(c)} \\
\hline
\multicolumn{4}{c}{dwarfs} \\
\hline
T$_{\rm eff}$  & 2.177$\pm$0.071 & 2.184$\pm$0.098 & 2.186$\pm$0.092 \\
$\leq 5700$\,K & 2.177$\pm$0.084 & 2.176$\pm$0.108 & 2.185$\pm$0.102 \\
\hline
T$_{\rm eff}$  & 2.215$\pm$0.074 & 2.243$\pm$0.076 & 2.235$\pm$0.075 \\
$\leq 6000$\,K & 2.220$\pm$0.074 & 2.257$\pm$0.075 & 2.253$\pm$0.074 \\
\hline
\multicolumn{4}{c}{turnoff and subgiant stars}\\
\hline
T$_{\rm eff}$ & 2.260$\pm$0.098 & 2.297$\pm$0.119 & 2.307$\pm$0.127 \\
$\leq 5700$\,K & 2.252$\pm$0.099 & 2.301$\pm$0.126 & 2.284$\pm$0.106 \\
\hline
T$_{\rm eff}$ & 2.235$\pm$0.077 & 2.260$\pm$0.105 & 2.289$\pm$0.129 \\
$\leq 6000$\,K & 2.235$\pm$0.077 & 2.260$\pm$0.105 & 2.253$\pm$0.085 \\
\hline
\end{tabular}
\end{center}
\end{table}

After a careful inspection of the literature in order to identify single- and
double-lined binaries (we mainly used the extensive surveys and listings by
Latham et al. 2002, Carney et al. 1994,2003) we then tested if any of our above
conclusions on the mean lithium abundance and on the dispersion may have been
affected by the inclusion of binary stars. This is especially important because
the literature-based EW values may have been derived at a time when the binary
nature of the star was still unknown. In order to get the most conservative
answer, we also checked the SIMBAD database for binary and variable stars, and
we included also the stars marked as ``binaries" in BM97. All the stars thus 
identified are marked with different symbols in Table~6 (2nd column, see the
legend at the bottom of the version of the table provided on-line) and
the situation can be visualized in Fig.~\ref{ALi_vsTeff_vsFesurH_duplicity}. 
Over the whole metallicity and effective temperature ranges sampled in this
paper, 27 (10, 3) stars from the {\it clean} sample are confirmed binaries
(suspected binaries, variable stars). The corresponding numbers for the {\it
ubvy} and {\it $\beta$} samples are 1 (0, 0) and 4 (1, 3). 

We compute the average value of A(Li)$_{NLTE}$ and the average
deviation of the observational errors for two cases : We first omit all 
the suspected and confirmed binaries, and then we exclude only the confirmed
binaries. In both cases, the stars reported as variable in SIMBAD are not
considered. 

The results are reported in Table~14 for the dwarf stars on one hand 
and for the turnoff and subgiant stars together on the other hand. 
In each case the first and second lines refer respectively to the values 
of the {\it clean} and {\it complete} samples. We also recall the values
discussed previously, which were obtained regardless of the duplicity or
variability of the stars (see column a).

First, we see that in all the considered cases, the absolute numbers for the
mean lithium abundance are always slightly higher (although fully compatible
within the errors) when binary and variable stars are excluded. Secondly, the
dispersion increases but only slightly. This confirms the early finding by Molaro (1991)
that known binaries do not exhibit lithium abundances significantly different
than the other stars and that they do not introduce significant scatter into
the plateau. Finally, the finding that plateau dwarf stars exhibit a lower
lithium mean value (although fully compatible within the quoted errors) than
their evolved counterparts is robust and resists the duplicity check. 

\section{Stars with extreme Li abundances}

\subsection{Plateau stars with Li upper limits}

\begin{table}
\label{tableul}
\caption{Sample stars with Li upper limits. The stars with a $*$ lie in the 
ranges in effective temperature and metallicity chosen to delimitate the Li plateau.
The others are either cooler or more metal-rich. The $\#$ indicates the stars
for which a {\it v}~sin{\it i} value higher than 4.5 km.sec$^{-1}$ could
be derived by Ryan et al. (2002) or Ryan \& Elliott (2004). See the text for more 
details.}
\footnotesize{
\parbox{220mm}{
\setlength{\tabcolsep}{0.15cm}
\begin{flushleft}
\begin{tabular}{ccccccccc}
\noalign{\smallskip}
\hline
\hline
\noalign{\smallskip}
 HIP & &  & \teff & Log L/L$_{\odot}$ & [Fe/H] & Li u.l. & single  & \\
     & &  &       &                   &        &         & /binary & \\
\noalign{\smallskip}
\hline
Dwarfs \\
\hline
72561 &$^*$&$\#$&  6388 & -0.12$\pm$0.30  & -1.66  & 1.639 & b \\
100682 &$^*$&$\#$&  6362 & -0.12$\pm$0.20  & -2.83  & 1.088 & b \\
67655   & &      &  5429 & -0.38$\pm$0.04  & -1.03  & 1.245 & s \\
67863    & &     &  5683 & -0.04$\pm$0.04  & -0.88  & 1.338 & s \\
\hline
Subgiants \\
\hline
81276 &$^*$&$\#$&  6523 &  0.76$\pm$0.58  & -1.50  & 1.705 & b \\
83320 &$^*$&$\#$&  5984 &  0.71$\pm$0.77  & -2.56  & 1.265 & s \\
55022  & &$\#$&  6367 &  0.50$\pm$0.14  & -1.29  & 1.313 & s \\
60719   & &      &  5319 &  1.44$\pm$0.21  & -2.32  & 0.789 & s \\
\noalign{\smallskip}
\hline
\end{tabular}
\end{flushleft}
}
}
\end{table}

In all the above discussions we have quoted the values for the 
mean lithium abundance and dispersion as well as for the trends with 
effective temperature and metallicity obtained without taking
into account the stars with Li upper limits. The main characteristics
of these Li-depleted stars are summarized in Table~15. 
In the metallicity and effective temperature ranges we have chosen to
delimit the plateau, we have two dwarfs (HIP 72561 and 100682) and two
more evolved stars (HIP 81276 and 83320). The first 3 of these objects
have relatively high effective temperatures and are either dwarfs or
subgiants lying very close to the turnoff, while the last one is clearly
crossing the Hertzsprung gap. 

Several studies have been devoted to these so-called ultra-lithium-deficient 
(hereafter ULDs) halo stars\footnote{Additional ULDs which can
be found in the literature do not have Hipparcos parallaxes and 
are thus excluded from our sample and from the present discussion. 
See references in Ryan et al.~(2002).}. 
Understanding their nature is crucial.  
They should indeed be excluded  from the investigations on the 
primordial Li abundance if they belong to a very special class of objects. 
However if they appear to be the extreme representatives of a process that
has affected all the stars along the plateau, they would be precious clues
on Li depletion in halo stars. 
Up to now, no mechanism has been unambiguously identified as responsible
for the ULD phenomenon. 
 
Norris et al.~(1997) looked at the abundances of many elements in the 
spectra of HIP 72561, 83320 and 100682 and found no common abundance 
anomaly that could be associated with the Li deficiency (see also
Ryan et al.~1998 and Ryan \& Elliot 2004). HIP 100682 
presents an overabundance of the heavy-neutron capture elements (a feature
which is shared by $\sim$ 25$\%$ of halo objects with similar [Fe/H])
whereas none of the other two ULDs shows such an enrichment.

Hobbs et al. (1991) and Thorburn (1992) suggested that HIP 83320 and 100682
might be the progeny of blue stragglers that had depleted their Li to
undetectable levels earlier on and are now evolving redwards, through the
temperature range of the plateau. However Thorburn noticed that there are
too many ULDs compared to the number of known halo blue stragglers. 

The possible role of duplicity in producing the observed Li depletion in the
ULDs via mass transfer has been discussed several times in the literature. 
Ryan et al. (2001a) proposed that the same mass transfer 
processes responsible for the field blue stragglers should also produce
sub-turnoff-mass objects which would be indistinguishable from normal stars
except for their Li abundance. This could in principle lead to Li depletion
and Ba enhancement as observed in HIP 100682 
if the donor companion was an AGB star. On the other hand, there also exist
the possibility that the mass transfer from an RGB or an AGB star is not
always accompanied by other abundance anomalies. HIP 72561
 and 81276 appear to be multiple systems and HIP 100682 is a suspected binary
(Carney et al. 1994, 2003, Norris et al. 1997). HIP 83320 however is a single
star. 

Ryan et al. (2002, hereafter R02) looked in detail at a sample of 
18 halo main-sequence turnoff stars including 4 ULDs\footnote{Two of 
these objects do not appear in our study, one of them (CD-31$^o$19466) 
because it does not have an Hipparcos parallax, the other because of 
its relatively high metallicity (BD+51$^o$1817, [Fe/H]~=~$-$0.88).}. 
They discovered that 3 out of these 4 Li-depleted stars (but none of the
Li-normal stars) exhibited unusually broad absorbtion lines that could be
attributed to rotational broadening. Among these objects are HIP 72561
and 81276 for which values of 5.5$\pm$0.6 and 8.3$\pm$0.4~km.sec$^{-1}$
were respectively inferred for {\it v}~sin{\it i} (Note that the ``Li-normal"
stars have undetectable rotation, generally below 3km.sec$^{-1}$). 
The 3 broadened objects of R02's sample all have relatively high effective
temperature and they lie close to the turnoff. Furthermore, they are all
confirmed binaries with orbits that are not tidally synchronised. These
complementary features lead R02 \footnote{This followed Fuhrmann \& Bernkopf
(1999) who suggested such a relation in the case of thick-disk, binary blue
stragglers.} to draw a connection between Li depletion, rapid rotation 
and mass and angular momentum transfer from a companion, now the white dwarf
remnant of a star initially more massive than the present plateau stars, as
confirmed by the projected companion masses inferred by Carney et al. (2001). 
Ryan \& Elliott (2004) looked again at line-broadening in ULDs and showed
that 5 out of 8 have rotation velocities in excess of 4km.sec$^{-1}$ and
that 4 out of 5 are confirmed binaries. The invoked mass transfer thus has 
not resulted in the merger of the components. 

In the scenario proposed by Ryan and collaborators, the Li depletion could be
due either to mixing triggered during the accretion event, or to Li deficiency
of the donor only or of both companions prior to the mass transfer. If
additional evidence could be found in favor of this scenario for the formation
of ULDs, it would become clear that these objects are not useful to infer the
primordial abundance nor to constrain the classical depletion mechanism(s).
However, although this explanation is appealing and may work for the
line-broadened and hot ULD stars, the unbroadened and/or single ULDs still
await a plausible interpretation.
As we shall discuss in more detail later on, the fact that all these
stars lie on or originate from the hottest side of the plateau may
reveal an alternative explanation. 

\subsection{Li upper limits outside the plateau range}

In our sample some other stars that are cooler and/or more metal-rich 
than the plateau also have Li upper limits, namely the dwarfs HIP 67655 
and 67863 and the post-turnoff stars HIP 55022 and 60719 
(see Table~15). 

HIP 55022 is a relatively hot and metal-rich subgiant that lies very close
to the turnoff. It is a confirmed binary, and could be the high-metallicity
counterpart of the ULDs previously described.  
Ryan \& Elliott (2004) give a {\it v}~sin{\it i} value of
10.4$\pm$0.2km.sec$^{-1}$ for this star. 

All the other 
three objects are single stars. HIP 67655 and 67863 are cool
dwarf stars that lie in the effective temperature
region where the lithium depletion relative to the plateau is strong 
(see \S~10.5). Therefore, the non-detection of Li in these 
two stars does not appear to be a real abnormalty. 

HIP 60719 is a single star at the base of the RGB that is undergoing Li
dilution. Although this could explain its Li upper limit, this star 
clearly lies below the other evolved stars of our sample indicating that it has
undergone a more severe lithium depletion. It is slightly more evolved than
the single subgiant HIP 83320 which was discussed previously. For none of
these two stars can mass transfer be advocated to explain the Li behavior. 
Note that HIP 60719 and 83320 additionally contribute to the Li dispersion 
observed in evolved stars and discussed in \S~11.2. 

\subsection{Stars with abnormally high Li abundances}

On the other extreme, some stars of our sample present relatively high
lithium abundances, namely HIP 86694 and 99423. Both of them
are post-turnoff stars.
We have tried to identify in our sample other stars that share the same
characteristics as these objects, but differ in their Li abundance. 

No real pair star could be attributed to HIP~86694. This star 
appears in several independent studies which all confirm its high Li content.
A(Li)$_{NLTE}$ values between 2.481 and 2.554 are obtained on the basis
of the extreme literature EW determinations. 

HIP~99423 has the same characteristics as HIP~73385 and 114962 
(Teff of 5806, 5831, 5883\,K; [Fe/H] of -1.54, -1.62, -1.52; 
LogL/L$_{\odot}$ of 0.482, 0.552, 0.554) but a significantly higher 
Li abundance (2.620, 2.383 and 2.245). Although HIP~99423 seems to be 
a single star, its ``pairs" are suspected or confirmed binaries. 

\subsection{Clues for another Li history in stars originating from the hot
side of the Plateau?}

We summarise our findings concerning the stars that present 
Li abnormalities, i.e., Li deficiencies or high Li values :

\begin{itemize}
\item Except for HIP 67655 and 67863 which are cool dwarf stars
lying in the effective temperature range where substantial lithium depletion
is expected to occur, all the other stars of our sample for which only an
upper limit for Li could be derived are either 
\begin{enumerate}
\item dwarf or turnoff stars at the extreme hot end of the plateau, or 
\item slightly more massive subgiants that have evolved from this
blue region.
\end{enumerate}

\item The stars with abnormally high Li are all post-turnoff stars. 
\end{itemize}

In other words, all 
these objects lie on or originate from the hot side of the plateau, 
and are thus more massive than the plateau dwarf stars for which 
no Li dispersion nor any other anomalies have been detected. 
This has to be related to the non-negligible lithium dispersion
and the higher lithium mean value that we get for post-main sequence stars 
compared to plateau dwarfs in a given range in effective temperature (see
\S~11). It is thus tempting to suggest that the most massive of the
halo stars still observable exhibit a Li dispersion together with some
``extreme" Li behavior (i.e., Li over-depletion or Li preservation) 
which reflect a different Li history to that of the less massive plateau
dwarf stars. We will come back to this crucial point in \S~14.

\subsection{Stars with $^6$Li detection}

Although we do not aim to homogeneize the $^6$Li/$^7$Li data available in
the literature, we need to comment on the very few halo stars in which $^6$Li
has been detected. This isotope is extremely difficult to observe, being only
a weak component, blending with the much stronger $^7$Li doublet at 670.7~nm.
The isotopic separation is 0.16 \AA~only. So far it has been 
detected in only a few stars which all belong to our sample, namely
HIP 8572 (G271-162), HIP 48152 (HD 84937) and HIP 96115 (BD + 26$^o$3578).
For all the other halo stars that have been looked at, only upper limits could
be derived (see Smith et al. 1993, 1998, 2001; Hobbs \& Thorburn 1994, 1997;
Cayrel et al. 1999a; Hobbs et al. 1999; Nissen et al. 2000). 

In Table~16 we recall the main characteristics of our sample stars for which
positive detections of $^6$Li have been reported and published. Note that the
3 stars under scrutiny are at the turnoff (i.e., with relatively high \teff~)
and have relatively low [Fe/H] values\footnote{Standard models of Pop II stars
predict only a slight depletion of $^6$Li (which occurs mainly during the
pre-main sequence phase) for stars which are now at the turnoff. The depletion
factor increases for lower mass dwarf stars
(see e.g. Deliyannis \& Malaney 1995 and Cayrel et al. 1999b). This explains
why $^6$Li has been found so far only in turnoff stars relatively metal-poor}.
They have no additional peculiarities except for $^6$Li when compared to other
stars in which this isotope has been looked for but not detected (see e.g.
Smith et al. 1998). 

\begin{table}[t]
\label{tableul}
\caption{Sample stars with published $^6$Li detection. 
The data are from Smith et al. (1998, sln98), Cayrel et al. (1999a, c99) and
Nissen et al. (2000, n00). Columns 2 and 3 give the \teff(2) and [Fe/H]
values that we attributed to the stars while columns 4 and 5 display the
values quoted in the original papers. For completeness, we note that a
re-analysis by Asplund et al. (2005) of HIP 8572 and HIP 96115 from very
high resolution and S/N UVES (Dekker et al. 1999) spectra finds a slightly
less than 2$\sigma$ detection for HIP 8572 and no $^6$Li detection for HIP
96115.}
\footnotesize{
\parbox{220mm}{
\setlength{\tabcolsep}{0.15cm}
\begin{flushleft}
\begin{tabular}{cccccccc}
\noalign{\smallskip}
\hline
\hline
\noalign{\smallskip}
 HIP   & \teff & [Fe/H] & \teff    & [Fe/H]   & Li      & $^6$Li/$^7$Li & Ref \\
       & (2)   &        & lit & lit & NLTE    &               &     \\
\noalign{\smallskip}
\hline
8572   & 6287 & -2.51 & 6295 & -2.15 & 2.236 & 0.02$\pm$0.01 & n00 \\
48152  & 6377 & -2.28 & 6300 & -2.30 & 2.249 & 0.05$\pm$0.02 & c99 \\
       &      &       & 6300 & -2.25 &       & 0.06$\pm$0.02 & n00 \\
96115  & 6322 & -2.42 & 6280 & -2.60 & 2.223 & 0.05$\pm$0.03 & sln98 \\
\noalign{\smallskip}
\hline
\end{tabular}
\end{flushleft}
}
}
\end{table}

Additionally, recent results from Asplund et al. (2005) indicate
nine new $^6$Li detections, in stars spanning a range in metallicity from
[Fe/H]=$-$1.3 down to $-$2.7. Four out of these nine objects belong to our
sample and again turn out to be stars with relatively high effective
temperatures, close to the turnoff, and with relatively low metallicity.

$^6$Li, together with $^9$Be, and $^{10}$B (the exact contribution from
$\nu$-spallation in supernovae to $^{11}$B being still under debate), is
believed to originate primarily (plus  some extra contribution by
$\alpha-\alpha$ reactions and other stellar sources, the latter important
at higher metallicites) from spallation reactions in the interstellar
medium between cosmic-ray (CR) $\alpha$-particles and protons and heavy
nuclei like CNO (Reeves et al. 1973) . Basic considerations about the
production rate and environment of these light nuclides predict a quadratic
slope in the logarithmic plane [(light nuclide, e.g. $^6$Li),metallicity],
whereas the almost linear slope derived from spectroscopic analyses of Be
and B abundances in stars of the Galactic halo (e.g. Boesgaard et al. 1999 
for Be, Primas et al. 1999 for B) seems instead
to indicate a primary (instead of secondary) origin, which in turn requires
the need for a production mechanism independent of the metallicity in the
ISM. One can generally refer to
the two abovementioned scenarios as to {\sl classical} {\it vs} {\sl reverse} 
spallation reactions, with the latter having several implementations based on
different assumptions.

Despite the small existing number of $^6$Li detections, this light nuclide
seems to have a different history, showing a much flatter evolution (i.e. most
of GCR-based models underproduce the amount of $^6$Li observed in 
halo stars). 

Because $^6$Li is the most fragile isotope to proton destruction,
its presence in the atmosphere of Pop II stars is usually considered as a
very severe limit on the amount of $^7$Li depletion. This argument is very
often used in favor of the Li plateau abundance being the primordial value
(Brown \& Schramm 1988), a conclusion that is challenged now by the CMB
constraint.
In this context several questions remain open. First, what is the
pre-stellar value of $^6$Li/$^7$Li (Probably not the one we observe now,
because of the different sensitivity of both isotopes to nuclear destruction)?
How can we explain the presence of $^6$Li in some of the plateau stars?
Why do some turnoff stars exhibit some $^6$Li in their spectra while the
majority does not? 

One possibility calls for some $^6$Li production at the stellar surface by
either stellar flares (Deliyannis \& Malaney 1995, but see the criticism by
Lemoine et al. 1997) or Galactic cosmic rays (Lambert 1995). The other
possibility is more related to the questions about the origin of $^6$Li in
metal-poor stars and about a possible scatter in the $^6$Li/$^7$Li ratio in
the ISM at a given metallicity. The type of cosmic ray sources and the
production mechanims operating in the early, forming galaxy are still very
controversial (see e.g. Vangioni-Flam et al. 1999). Some models for the
formation of light elements by cosmic ray processes predict a scatter of one
order of magnitude in the abundances of $^6$Li, Be and B relative to Fe. This
is the case of the bimodal superbubble model by Parizot \& Drury (1999) and
of the supernovae-driven chemical evolution model for the Galactic halo by
Suzuki et al. (1999). 
The superbubble scenario is actually challenged by several observations, 
for example  the fact that there seems to exist no association of core
collapse SN with superbubbles but rather to HII regions which reflect the
metallicity of the ambiant interstellar medium rather than that of the SN
(see the discussion in Prantzos 2004).

Suzuki \& Inoue (2004) recently presented an additional mechanism for cosmic
ray production of $^6$Li by virialisation shocks during hierarchical structure
formation of the Galaxy. In their model, cosmic rays accelerated by this
source dominate the production of $^6$Li (compared to SNe as the main source
of acceleration), without co-producing Be and B. This seems to better account
for the observed $^6$Li data in halo stars,
and in particular for the ``plateau" of $^6$Li/H reported by Asplund et al.
(2005), than the SN-driven cosmic ray scenarios. Large $^6$Li scatter should
be a natural consequence of this process, a higher initial content being
expected  in stars belonging to the inner halo compared to those of the
outer halo. This is an attractive scenario, although very difficult to
estimate quantitatively due to the many uncertainties on the physics of
structure formation and on the energetics of the implied cosmic rays.

Clearly the detection of $^6$Li in low metallicity halo stars constitutes a
challenge to cosmic ray spallation and/or stellar mixing 
along the plateau. Alternatively it may be the case that the
lithium isotope ratio probes Physics beyond the Standard Model. 
Indeed Jedamzik (2004) found that the decay of a supersymetric 
particle like the gravitino or neutralino around 10$^3$ s after the Big Bang 
may at the same time yield to a reduction of the primordial $^7$Li/H by a factor 2-3 
and produce $^6$Li to the magnitude observed in halo stars. However before
we turn to such possibilities, the astrophysical uncertainties and 
solutions must be critically assessed.

\section{Implications for stellar structure and evolution}

Since the discovery of lithium in Pop II stars, not only the observers have
been very active on the subject. Theoreticians, too, have taken advantage of
the lithium diagnostic to probe the interior and the evolution of Pop II
stars. 
Although these low-mass objects seem to be very simple at first sight, 
their Li behavior is paradoxal and has still not been fully understood. 
Part of the difficulty comes from the fact that nothing is known about the 
initial conditions of an important quantity like the 
angular momentum distribution. 
We do not aim here to discuss all the literature published 
on the subject (see the review by Pinsonneault et al. 2000 and 
Talon \& Charbonnel 2004 for more recent references). 
We rather use the specific findings of our analysis to propose a synthesis, 
i.e., to extract from the theoretical debate some of the most adapted and 
promising directions that require further investigation. 

The (incorrectly) so-called ``standard" stellar models\footnote{This 
refers to the modeling of non-rotating, non-magnetic stars in which
convection is the only transport process considered} have long been 
very popular in the plateau debate, mainly because they predict 
no variation of the surface Li abundance with time and consequently 
no Li dispersion from star to star along the plateau. 
The corresponding Li isochrones thus naturally support the conclusion that 
the Spite plateau reflects the cosmological Li abundance. 
This belief has been shaken by the recent CMB measurements.

In addition to and maybe even more fundamental 
(from the stellar physicist point of view) 
than the CMB result, the validity of these stellar models can be refuted 
on the basis of some of their assumptions, the most critical of which being
the absence of transport of chemicals except in the convection zones. 
This hypothesis denies the fundamental nature of a star which 
is a gaseous mixture of elements with various atomic masses. 
As a result the stellar gas cannot be in equilibrium as a whole until 
each component reaches its own equilibrium via gravitational
separation and thermal diffusion (Eddington 1929).
Thanks to helioseismology it is now fully recognised that atomic diffusion 
must be an integral part of stellar 
evolution computations and in particular of the {\it standard}
model\footnote{It is now widely accepted that the {\it standard} stellar models 
are those in which the effects of atomic diffusion are taken into account and not 
counterbalanced by any macroscopic process. They are calculated from first principles 
without any arbitrary parameter except for the mixing length.}
of the Sun (i.e., Richard et al. 1996, Bahcall et al. 1997) and of low-mass
stars.

The relative unpopularity of the Pop II {\it standard} models rests mainly 
on the fact that pure atomic diffusion leads to a degree of surface Li
depletion which increases with effective temperature along the plateau
(Michaud et al.~1984; Deliyannis et al.~1990; Proffitt \& Michaud 1991;
Chaboyer \& Demarque 1994; Vauclair \& Charbonnel 1995; Salaris \& Weiss 2001;
Richard et al.~2002). As confirmed again in the present analysis, this feature
is not observed, although some of the hottest dwarf stars with Li deficiency
may exhibit the effects of atomic diffusion (see \S~13.1 and discussion
below). This difficulty can be simply read as the signature of 
some macroscopic processes that minimize 
the effects of atomic diffusion which is always present in stellar interiors
and cannot be arbitrarily turned off. Such a suggestion was already made
in a more general context by Eddington (1929) who pointed out that some mixing
must occur in the stellar radiative zone in order to prevent the gravitational
settling and the thermal diffusion of heavy elements, the effects of which were
not always observed.
Consistently, such a process is also required in Pop II stars 
to counteract the settling of heavier elements in order to explain
the close similarity of iron abundances in near turnoff, sub-giant
and lower RGB stars in globular clusters, and to reproduce the
observed morphologies of globular cluster color-magnitude
diagrams (see VandenBerg et al. 2002).

Several processes have been invoked to counteract atomic diffusion in stellar 
interiors. In the case of Pop II stars the possible candidates that have
been studied (in models including or not atomic diffusion) are : 
Rotation (Vauclair 1988; Chaboyer \& Demarque 1994; Pinsonneault et al. 1991,
1999, 2002; Vauclair \& Charbonnel 1995), stellar wind (Vauclair \& Charbonnel
1995), interaction between meridional circulation and helium settling
(Th\'eado \& Vauclair 2001). All these models have real difficulties, unless
some adhoc assumptions are made, in reconciling a non-negligible Li destruction
with both the flatness and the extremely small dispersion on the plateau 
as definitively needed in view of the difference between the WMAP constraint
and the plateau value discussed in this paper. This does not mean of course
that the physical processes invoked are not at work in Pop II stars but it
indicates at least that their theoretical description is still incorrect or
incomplete. 

Richard et al. (2002, hereafter Ri02; 2004) re-investigated the case 
of the Li plateau with a new generation of Pop II models that include 
self-consistently all the effects of atomic diffusion in the presence of 
weak turbulence whose nature is not postulated a priori. 
They discuss in great detail the parametrization of turbulence that has 
to be included in order to 
reproduce the constancy of the Li plateau (see also Proffitt \& Michaud 1991 
and Vauclair \& Charbonnel 1998 and references therein). 
The mean Li value we derive in the present paper for the plateau dwarf stars
favors such a model with an intermediate efficiency of turbulence (actually
between the so-called T6.0 and T6.25 models of Ri02~\footnote{In a Tx.y model,
the turbulent diffusion coefficient is 400 times larger than the He atomic
diffusion at log T~=~x.y and varies as $\rho^{-3}$.}).
In addition Ri02 show that acceptable variations in the
turbulence (i.e., from no macroscopic motion to that needed to fit the
plateau) can lead to variations of the Li abundance as high as 0.5~-~0.6~dex
at the turnoff. As can be seen in their Figs.~14 and 16 the Li abundance at
that evolutionary stage is indeed a very sensitive function of the exact
position in the HR diagram and of the adopted turbulence. 
The abundance variations and in particular the Li deficiencies that we find 
in our data at or just after the turnoff as well as those found by B98 
in M92 are similar to those expected at this evolutionary stage by Ri02 in 
the case of variations of turbulence from star to star. 
However in Ri02's models these abundance variations are 
theoretically erased by dilution in the subgiants when they reach $\sim$~6000K,
 although the dispersion persists in the data at and below this effective
temperature. 
This difficulty may be alleviated by assuming that some stars 
have undergone even stronger turbulence that lead to stronger Li destruction. 

As previously mentioned, Richard and collaborators do not postulate 
the physical mechanism that causes the turbulence required by their models. 
Our present results bring a very important piece to the puzzle. 
We could show that the turnoff and more evolved
Pop II stars present a slightly higher Li mean value as well as 
a larger Li dispersion than the 
less massive dwarfs. 
This points towards a mechanism, the efficiency of which changes as one 
reaches the extreme blue edge of the Li plateau.
Such behavior corresponds to that of the generation and filtering of 
internal gravity waves in Pop II stars. As shown by Talon \& Charbonnel (2004) 
gravity waves are indeed very efficient in dwarf stars along the 
plateau up to an effective temperature of $\sim$~6300\,K. 
There they dominate the transport of angular momentum 
and should lead to a quasi-solid rotation state of the stellar radiative zones 
on very short time-scales. 
As a result the surface Li depletion is expected to be independant of the
initial angular momentum distribution in this range of effective temperatures. 
This should alleviate the difficulty encountered by the classical 
rotating models which predict that a range of initial angular momenta 
generates a range of Li depletion and that the scatter increases 
with the average Li depletion. 
In more massive stars however the efficiency of the gravity waves
strongly decreases and 
internal differential rotation is expected to be
maintained under the effect of meridional circulation and turbulence 
induced by rotation\footnote{The mass dependance of the gravity waves efficiency 
leads to a natural explanation of the fast horizontal branch rotators. 
It also provides a solution to the enigma of the so-called Li dip observed
in Pop I stars in terms of rotational mixing, forming a coherent picture 
of mixing in main sequence stars of all masses. Also, gravity waves 
are able to shape the Sun's flat rotation profile deduced from helioseismology.
See Talon \& Charbonnel (2003, 2004) and Charbonnel \& Talon (2005) for more details.}. 
Consequently variations of the original angular momentum from star to star 
would lead to more Li dispersion and to more frequent abnormalities 
in the case of the most massive stars where gravity waves are not fully efficient. 
In other words internal gravity waves are expected to dominate completely 
the transport of angular momentum and should lead to higher Li homogeneity 
in the less massive, rigid rotators than in the stars lying on or originating 
from the hot side of the plateau. 
The proper treatment of the effects of gravity waves together with those 
of atomic diffusion, meridional circulation and shear turbulence has
now to be undertaken in stellar evolution models in order to test the real 
and interactive consequences of all these complex mechanisms 
(Talon \& Charbonnel 2005).

\section{Summary and conclusions}
The Spite \& Spite (1982a,b) discovery has set the stage for 
analyses to follow focusing on the determination of the lithium abundances 
in the most metal-poor, thus the oldest stars of our Galaxy.
In view of the crucial importance of this problem for cosmological,
galactic and stellar implications, all the observational and theoretical 
aspects have been the subject of very lively debates. In the present paper
we tackled further the most critical issues by revisiting
the Li data available in the literature. Our sample was assembled following
strict selection criteria on the quality of the original analysis, i.e.,
high resolution and high signal to noise spectra. 

In the first part we focused on the systematic uncertainties affecting the
determination of Li abundances. We explored in detail the temperature
scale issue and put special emphasis on reddening with the aim of deriving a
tool as consistent as possible for all our sample stars. In order to do so,
we chose to derive photometric temperatures using Str\"omgren {\it uvby-$\beta$}
photometry which was the only one available for our entire sample. During
these steps, we identified one of the major drawbacks of such determinations,
namely an accurate estimate of the reddening affecting each of our stars.
We derived four sets of effective temperatures based on different assumptions 
for the interstellar reddening excess values, E({\it b-y}).
We tried to evaluate the effect of using reddening values taken from different
sources on the derived temperature scales and in turn on the derived A(Li)
abundance, showing that an unpredictable mix of different reddening sources could
be held responsible for opposite findings, on the same dataset, about the presence
of dispersion and/or slope on the Li plateau.   
Finally, we selected as our best and final \teff~scale the one (\teff(2)) that has been
derived from de-reddening all the stars, except those with negative E({\it b-y})
values. 

In order to keep as many stars as possible in our analysis we had to make some
compromises on the derivation of the effective temperature of some objects. 
This contaminated our {\it complete} sample of 
118 stars which was finally subdivided as follows : 
1) The {\it clean} sample which contains 91 stars for which
the complete set of Str\"omgren photometric indices are available 
and for which the Schuster \& Nissen (1989) calibration
for the interstellar reddening excess is applicable.
2) The {\it $\beta$} sample which includes 
20 stars for which the reddening E({\it b-y}) value was derived from 
averaging different sources of E(B-V) via the Crawford's formula.
3) The {\it ubvy} sample which contains 7 stars for which one of the
{\it ubvy} photometric indices was found to fall just slightly outside
the allowed intervals for the application of the Schuster \& Nissen 
calibrations. 
We made several tests to quantify the influence of these compromises 
on the statistical analysis and on our final assessments.
In order to guarantee the absence of spurious differences and 
conclusions due to the use of different criteria in the determination 
of the effective temperature, 
all our results (summarised hereafter) regarding the characteristics
of the plateau were given for the {\it clean} sample on one hand and
for the {\it complete} (i.e., {\it clean + $\beta$ + ubvy}) sample on
the other hand.
This approach should provide as high accuracy and reliability as possible 
for one of the largest sample yet studied. 

We then derived the lithium abundances for the various subsamples using 
the different sets of temperature. This was done using the arithmetic 
mean of the equivalent widths and the 1~$\sigma$ uncertainty of the 670.7~nm 
line as reported in the literature from which we had assembled the sample. 
The lithium abundance was first derived under LTE assumptions for all the
\teff~scales studied here and then NLTE corrections were applied. With these
NLTE Li abundances we determined the mean Li value and dispersion 
along the plateau for our sample as a whole. 
In order to avoid any contamination by lithium production from various 
stellar sources we then restricted our discussion only to those stars 
with [Fe/H] lower than $-$1.5. In this metallicity range, we considered as 
plateau stars those with an effective temperature higher than 5700 or 6000\,K.
Stars with Li upper limits were excluded from the analysis and are discussed 
separately later. 
We found sligthly different results for the mean lithium abundance 
and for the dispersion depending on the lowest limit on \teff 
and on the sample under consideration, i.e. {\it clean} or {\it complete}.
We note however that the pollution due to our compromises on the 
derivation of the effective temperature of some of our objects has a
negligible impact on our conclusions. The effect of having assumed 
a lower limit on [Fe/H] in the colour-\teff\,calibration
used here to derive our sets of effective temperatures is also negligible. 
This assumption has only the effect of raising the A(Li) plateau
abundances by 0.03~dex, on average, making our conclusions quite robust. 

In the case of the stars of the {\it clean} sample with \teff~$\geq$~6000\,K 
we obtain $$A(Li)_{NLTE} = 2.2243 \pm 0.0748 .$$ 
This is a factor of 2.48 to 2.74 lower (depending on the SBBN study we rely
on,  i.e., Coc et al. 2004, Cyburt 2004 or Serpico et al. 2004) than the 
prediction for a standard Big Bang corresponding to the WMAP estimate of
$\Omega_b h^2$. The relatively low lithium abundance 
seen in metal-poor halo stars is a very robust result.  
Assuming the correctness of the CMB constraint on the value of the
baryon-to-photon ratio we are then left with the conclusion that the Li
abundance seen at the surface of halo stars is not the pristine one, but
that these stars have undergone surface lithium depletion at some point
during their evolution.

In the second part of the present paper we further pushed the constraints on 
lithium depletion in halo stars. 
Using our homogeneized data we looked at the Li plateau by considering 
the evolutionary status of each star. 
This could be done using the Hipparcos parallaxes which were
available for almost all our initial sample stars. 
This step of the analysis proved to be crucial since a contamination 
exists from post-main sequence stars, which has to be removed in order 
to precisely determine the depletion factor along the plateau. 
Several conclusions could then be drawn. 

Again the mean lithium abundance for the dwarf stars depends on 
the lowest effective temperature chosen to delimit the plateau and 
slightly varies when one considers the {\it clean} sample only or 
the {\it complete} one. 
The mean lithium plateau value for the dwarf stars of
the {\it clean} sample with \teff~$\geq$~6000\,K is 
$$A(Li)_{NLTE} = 2.2154 \pm 0.0737.$$  
This is a factor of 2.53 to 2.8 lower than the WMAP + SBBN primordial Li value.
Note that for the dwarf stars of the {\it clean} sample with
\teff~$\geq$~5700\,K we derive a 
mean value of 
$$A(Li)_{NLTE} = 2.1768 \pm 0.0711$$ 
which is 2.76 to 3.06 times lower than the CMB-derived value. 
We find no evidence of intrinsic Li dispersion along the plateau
when only the dwarf stars are considered. 

A very surprising result was found for the first time : 
Whatever the subsample we considered, the mean value of A(Li)$_{NLTE}$
always appears to be higher (although compatible within the errors) 
for the subgiant stars than for the dwarfs, except for the most metal-deficient 
objects (i.e., with [Fe/H]$\leq$-2.5) where the mean lithium abundance is very 
similar in both evolutionary status. 
The mean lithium value for the post-main sequence stars of
the {\it clean} sample with \teff~$\geq$~6000\,K is
$$A(Li)_{NLTE} = 2.2349 \pm 0.0769 $$
and with \teff~$\geq$~5700\,K is
$$A(Li)_{NLTE} = 2.2599 \pm 0.0997.$$
Additionally the post-main sequence stars show a non-negligible 
Li dispersion. This is true at the turnoff and all along the
Hertzsprung gap. This feature recalls that observed in subgiant stars
of the M92 globular cluster. 

We checked that the above conclusions do not change when we exclude 
the stars that belong to multiple systems or show variability. We could confirm 
that binary stars do not exhibit lithium abundances significantly different
to their single counterparts. 

We finished our close examination of our sample by looking at the stars that 
present deviant Li abundances, i.e., the stars with strong Li deficiency 
(the so-called ULDs) and those with an abnormally high Li content. 
We found that all of them lie on or originate from the hot side of the plateau. 
This agrees with our finding that the turnoff and subgiant stars 
present a slightly higher Li mean value and dispersion than the dwarfs. 
These results indicate that the post-main sequence halo stars experienced 
a Li history slightly different from that of the less massive plateau dwarfs.
We suggest that such a behavior may be the signature of a transport process of the chemicals 
and of angular momentum whose efficiency changes on the blue edge of the plateau.
This in agreement with the fact that most of the ULDs are presently rotating 
faster than the Li-normal stars. 
Our analysis provided thus some crucial clues to the internal processes 
that may be involved in modifying the surface Li abundances in halo stars. 
Since internal gravity waves coupled with rotation-induced mixing are 
expected to lead to higher Li homogeneity with \teff~ in the plateau 
stars than in the more massive stars lying on or originating from the
hot side of the plateau, such a model is favoured. 
Although we excluded the ULDs stars from the analysis and focused on
the ``normal-Li" stars to derive the Li mean values and the trends 
with \teff~ and metallicity, these objects should not be 
excluded from hydrodynamical studies of the Li depletion mechanisms 
that affect the Pop II stars. 

\acknowledgements{We dedicate this paper to the two new little 
lithium boys Angel David and Martin Lou. 
F.P. is indebted to CNRS and the University Paul Sabatier 
in Toulouse for the Visiting Researcher contracts awarded.
F.P. warmly thanks the Laboratoire d'Astrophysique de
Toulouse et Tarbes, France for the warm hospitality 
received during her extended and frequent visits. C.C. thanks the french
Programme National de Physique Stellaire and Programme National Galaxies for
financial support. Part of this work has been carried out at the University of
Washington (Seattle, USA) where both F.P and C.C. participated in the INT
Programme on Nucleosynthesis (April 2002). 
We used the SIMBAD data base operated at the CDS (Strasbourg, France).}

{}

\end{document}